\documentclass[times,sort&compress,3p]{elsarticle}
\journal{Journal of Multivariate Analysis}
\usepackage[labelfont=bf]{caption}

\usepackage{amsmath,amsfonts,amssymb,amsthm,booktabs,color,epsfig,graphicx,hyperref,url}
\usepackage{bm}
\usepackage{colortbl}
\usepackage{subcaption}
\usepackage{xr}

\newcommand{\br}{\bm{r}}

\newcommand{\bx}{\bm{x}}

\newcommand{\bA}{\bm{A}}
\newcommand{\bB}{\bm{B}}
\newcommand{\bH}{\bm{H}}
\newcommand{\bI}{\bm{I}}
\newcommand{\bJ}{\bm{J}}
\newcommand{\bP}{\bm{P}}
\newcommand{\bR}{\bm{R}}
\newcommand{\bS}{\bm{S}}

\newcommand{\bX}{\bm{X}}
\newcommand{\bY}{\bm{Y}}

\newcommand{\bbeta}{\bm{\beta}}
\newcommand{\bmu}{\bm{\mu}}

\newcommand{\bkappa}{\bm{\kappa}}

\newcommand{\bgamma}{\bm{\gamma}}
\newcommand{\bSigma}{\bm{\Sigma}}

\definecolor{Gray}{gray}{0.9}

\newtheorem{theorem}{Theorem}
\newtheorem{corollary}{Corollary}
\newtheorem{remark}{Remark}
\newtheorem{example}{Example}

\begin{document}

\begin{frontmatter}

\title{Distributed Simultaneous Inference in Generalized Linear Models via Confidence Distribution}

\author[A1]{Lu Tang}
\author[A2]{Ling Zhou}
\author[A3]{Peter X.-K. Song\corref{mycorrespondingauthor}}

\address[A1]{Department of Biostatistics, University of Pittsburgh, Pittsburgh, PA 15261, USA}
\address[A2]{Center of Statistical Research, Southwestern University of Finance and Economics, Chengdu, Sichuan, China}
\address[A3]{Department of Biostatistics, University of Michigan, Ann Arbor, MI 48109, USA}

\cortext[mycorrespondingauthor]{Corresponding author. Email address: \url{pxsong@umich.edu}}

\noindent This paper is published in Journal of Multivariate Analysis, \url{https://doi.org/10.1016/j.jmva.2019.104567}. 
\vspace{0.5cm}

\noindent \textbf{For citation, please use} \\
\noindent
Tang, L., Zhou, L., \& Song, P.X.K. (2020). Distributed simultaneous inference in generalized linear models via confidence distribution. \textit{Journal of Multivariate Analysis}, 176, 104567.

\ \\

\begin{abstract}
We propose a distributed method for simultaneous inference for datasets with sample size much larger than the number of covariates, i.e., $N \gg p$, in the generalized linear models framework.
When such datasets are too big to be analyzed entirely by a single centralized computer, or when datasets are already stored in distributed database systems, the strategy of divide-and-combine has been the method of choice for scalability.
Due to partition, the sub-dataset sample sizes may be uneven and some possibly close to $p$, which calls for regularization techniques to improve numerical stability.
However, there is a lack of clear theoretical justification and practical guidelines to combine results obtained from separate regularized estimators, especially when the final objective is simultaneous inference for a group of regression parameters.
In this paper, we develop a strategy to combine bias-corrected lasso-type estimates by using confidence distributions.
We show that the resulting combined estimator achieves the same estimation efficiency as that of the maximum likelihood estimator using the centralized data.
As demonstrated by simulated and real data examples, our divide-and-combine method yields nearly identical inference as the centralized benchmark.
\end{abstract}


\end{frontmatter}

\section{Introduction}\label{sec:intro}

We consider simultaneous inference for the generalized linear model (GLM) under the situation where data are stored on distributed computer clusters instead of a centralized location. The use of distributed storage can be due to either large data volume or protection of individual-level sensitive data from leaving data-owning entities. Such distributed data presents great challenges in statistical analyses because the entire dataset cannot be loaded once to a single processor for computation \citep{fan2014challenges}. In the advent of cloud storage and computing, the method of divide-and-combine, also known as divide-and-conquer \citep{aho1974design}, has become the state-of-the-art in big data analytics to effectively improve scalability.
Divide-and-combine is a computational procedure that divides the data into relatively independent, smaller and computable batches, processes them in parallel and combines the separate results. However, not all existing statistical methods are directly parallelizable.
Some complicated methods require special treatment in order to be adapted to the parallel computing architecture; see for examples, parallel matrix factorization by randomized matrix approximation in \citep{mackey2011divide}, scalable bootstrap by bag of little boostraps in \citep{kleiner2014scalable}, divide-and-combine-type kernel ridge regression in \citep{zhang2015divide}, and communication efficient lasso regression in \citep{lee2017communication},
among others.
In this paper, we consider simultaneous inference for the GLM using divide-and-combine. While both the sample size $N$ and the number of covariates $p$ may be large in practice, here we focus on the case when $N \gg p$, but $p$ is not small and can vary from hundreds to thousands. The $N$ observations are split into $K$ mutually independent sub-datasets.

Meta-analysis is an example of divide-and-combine that combines summary statistics from independent studies, see for examples \citep{sutton2008recent, stangl2000meta, hedges2014statistical}.
The classical fixed-effect meta-analysis uses inverse variance weighted average to combine separate point estimates. Raw data can be processed locally and only summary quantities are communicated between machines to reduce cost of data transfer \citep{lee2017communication}.
In the development of distributed algorithms for statistical inference, one question arises naturally: are the proposed divide-and-combine estimators and the maximum likelihood estimator (MLE) obtained from the centralized data asymptotically equivalent, leading to comparable statistical inferences?
\citet{lin2010relative} showed that such a meta-estimator asymptotically achieves the Fisher's efficiency; in other words, it follows asymptotically the same distribution as the centralized MLE.
The Fisher's efficiency has also been established for a combined estimator by \citep{lin2011aggregated} through aggregating estimating equations under a relatively strong condition that $K$ is of order $O(n^{r})$ where $r < 1/3$ and $n$ is the sample size of a sub-dataset. Recently, \citet{battey2018distributed} proposed test statistics and point estimators in the context of the divide-and-combine, where the method of hypothesis testing is only developed for low dimensional parameters, and the combined estimator takes a simple form of an arithmetic average over sub-datasets. Different from \citep{battey2018distributed}, we consider simultaneous inference for all parameters and use the inverse of variance-covariance matrices to combine estimates.

Although the overall sample size is large, it is reduced $K$ times in the sub-datasets due to data partition. The sample size reduction and potentially unbalanced sample sizes across sub-datasets may cause numerical instability in the search for the MLE, especially in overfitted models when most of covariates are unimportant among all covariates that are included in the analysis.  As shown in Section~\ref{sec:simulation}, coverage probabilities of confidence intervals obtained by the classical meta-analysis method deviate drastically from the nominal level as $K$ increases.
This motivates the use of regularized regression to overcome such numerical instability. For regularized estimators, such as lasso \citep{tibshirani1996regression} and SCAD \citep{fan2001variable}, constructing confidence intervals is analytically challenging because: (i) sparse estimators usually do not have a tractable limiting distribution, and (ii) the oracle property \citep{fan2001variable} relying on knowledge of the truly non-zero parameters is not applicable to statistical inference since the oracle is unknown in practice.

When penalized regression is applied on each sub-dataset, variable selection procedures will choose different sets of important covariates by different tuning schemes. Such misaligned selection prohibits any weighting approaches from combining the separate results; both dimensionality and meaning of the estimates across sub-datasets may be very different. \citet{chen2014split} proposed a majority-voting method to combine the estimates of the covariates most frequently identified by the lasso across the sub-datasets. Unfortunately, this method does not provide inference for the combined estimator, and it is sensitive to the choice of inclusion criterion.
To fill in this gap, we propose a new approach along the lines of the post-selection inference developed for the penalized estimator by \citep{van2014asymptotically} and \citep{zhang2014confidence}, which allows us to combine bias-corrected lasso estimators obtained from sub-datasets.

In this paper, we use the confidence distribution approach \citep{xie2013confidence} to combine results from the separate analyses of sub-datasets. The confidence distribution, originally proposed by \citet{fisher1956statistical} and later formally formulated by \citet{efron1993bayes}, has recently attracted renewed attention in the statistical literature; see for examples, \citep{singh2005combining,xie2013confidence} and references therein. An advantage of the confidence distribution approach is that it provides a unified framework for combining distributions of estimators, so statistical inference with the combined estimator can be established in a straightforward and mathematically  rigorous fashion. Specifically related to divide-and-combine, \citet{xie2011confidence} developed a robust meta-analysis-type approach through confidence distribution, and \citet{liu2015multivariate} proposed to combine the confidence distribution functions in the same way as combining likelihood functions for inference, and showed their estimator achieves the Fisher's efficiency.
The step of combining via confidence distribution theory requires well-defined asymptotic joint distributions of all model parameters of interest, which, in the current literature, are only available for $p$ less than $n$, the sample size of one sub-dataset under equal data split. Here, we consider the scenarios where $p$ and $K$ can both diverge to infinity with rates slower than $N$.
Our new contribution is two-fold: (i) the combined estimator achieves asymptotically the Fisher's efficiency; that is, it is asymptotically as efficient as the MLE obtained from the direct analysis on the full data; and (ii) the distributed procedure is scalable and parallelizable to address very large sample sizes through easy and fast parallel algorithmic implementation. The latter presents a desirable numerical recipe to handle the case when the centralized data analysis is time consuming and CPU demanding, or even numerically prohibitive.

This paper is organized as follows. Section~\ref{sec:distributed} focuses on the asymptotics of the bias-corrected lasso estimator in sub-datasets. Section~\ref{sec:dividecombine} presents the confidence distribution method to combine results from multiple regularized regressions. Section~\ref{sec:simulation} provides extensive simulation results, and Section~\ref{sec:Application} illustrates our method by a real data. We conclude in Section~\ref{sec:discuss}. We provide key technical details in the Appendix and defer complete proofs and supporting information to the Supplementary Material.

\section{Distributed Penalized Regressions for Sub-datasets}\label{sec:distributed}

For GLM, the systematic component is specified by the mean of a response $y_i$ that is related to a $p$-dimensional vector of covariates $\bx_i$ by a known monotonic canonical link function $g(\cdot)$ in the form $\mu_i = E(y_i) = g^{-1}(\bx_i^{T}\bbeta)$, for subject $i\in \{1, \dots, N\}$. The random component is specified by the conditional density of $\bY = (y_1, \cdots, y_N)^{T}$ given $\bX = (\bx_1, \cdots, \bx_N)^{T}$. The variance of the response takes the form of $\operatorname{var}(y_i) = \phi v(\mu_i)$ where $\phi$ is the dispersion parameter and $v(\cdot)$ is the unit variance function \citep{mccullagh1989generalized}. The associated likelihood function is given by $\mathcal{L}_N(\bbeta; \bY, \bX) = \prod_{i=1}^N\exp[\{y_i\theta_i - b(\theta_i)\} / \phi + c(y_i, \phi)],$ where $\dot{b}(\cdot)=g^{-1}(\cdot)$ and the canonical parameters have the form $\theta_i = \bx_i^{T}\bbeta$, with $\bbeta$ being the $p$-element vector of regression parameters of interest.

The centralized MLE solution, $\hat{\bbeta} = \arg\max_{\bbeta} \mathcal{L}_{N}(\bbeta; \bY, \bX) $, in general has no closed-form expression, except for the Gaussian linear model, and is often obtained numerically by certain iterative algorithms such as Newton-Raphson.
Thus, it is not trivial to establish exact parallel algorithms that only require a single passing of each sub-dataset, and still achieve the same efficiency as the centralized MLE.
Sample partition naturally results in $K$ sub-datasets, each with size $n_k$, and $\sum_{k=1}^K n_k = N$.

This section focuses on deriving the regularized estimator and confidence distribution for a \textit{single sub-dataset} of sample size $n_k$ for a specific $k$. Since the method in this section is general to all sub-datasets, for ease of exposition, we suppress $k$ unless otherwise noted. We start by deriving the asymptotic properties of lasso regularized regression, as our divide-and-combine procedure is dependent on the asymptotic results. The regularization plays an important role in stabilizing numerical performance on the divided datasets, which will be shown in later sections. We use lasso \citep{tibshirani1996regression} in the development of this paper. With little effort, other types of regularization, such as SCAD \citep{fan2001variable} or elastic net \cite{zou2005regularization}, may be adopted in our proposed procedure.

\subsection{Lasso in Generalized Linear Models}
The lasso estimator is obtained by maximizing the following penalized log-likelihood function with respect to the regression parameters $\bbeta$ subject to a normalizing constant,
\begin{eqnarray*} \label{eqn:plkhd}
	PL(\bbeta; \bY, \bX) &\overset{\text{def}}{=}& \frac{1}{n}\mathcal{L}_n(\bbeta; \bY, \bX) - \lambda\|\bbeta\|_1 \\
	&\propto&  \frac{1}{n\phi} \sum_{i=1}^n\left\{y_{i}\bx_i^T\bbeta - b(\bx_i^T\bbeta)\right\} - \lambda\|\bbeta\|_1,
\end{eqnarray*}
where $\lambda$ is a nonnegative tuning parameter, and $\|\bbeta\|_1 = \sum_{j=1}^{p}\vert\bbeta_j\vert$ is the $\ell_1$-norm of the regression coefficient vector $\bbeta = (\beta_1, \cdots, \beta_p)^{T}$. Let $\hat{\bbeta}_\lambda = \arg\max_{\bbeta} PL(\bbeta;\bY,\bX)$ be a lasso estimator of $\bbeta$ at a given tuning parameter $\lambda\ge0$. Solution $\hat{\bbeta}_\lambda$ may be obtained by coordinate descent via \citet{donoho1994ideal}'s soft-thresholding approach, with the tuning parameter being determined by, {\em say}, cross-validation \citep{shao2012estimation}.

\subsection{Confidence Distribution for Bias-corrected Lasso Estimator}
To combine multiple lasso estimators obtained from separate sub-datasets, we need to overcome the issue of misalignment: the sets of selected covariates with non-zero estimates in the model are different across sub-datasets. Our solution is based on bias-corrected lasso estimators. The bias correction enables us not only to obtain non-zero estimates of all regression coefficients, but also, more importantly, to establish the joint distribution of regularized estimators. The latter is critical for us to utilize the confidence distribution to combine estimators, which will be described in Section~\ref{sec:dividecombine}.

Denote the score function by $\bS_n(\bbeta) = \frac{1}{n\phi}\sum_{i=1}^n\left\{y_i - g^{-1}(\bx_i^{T}\bbeta)\right\}\bx_i$. It is known that the lasso estimator, $\hat{\bbeta}_\lambda$, satisfies the following Karush-Kuhn-Tucker (KKT) condition: $ \textstyle
\bS_n(\hat{\bbeta}_\lambda) - \lambda \hat{\bkappa} = 0,$ where subdifferentials $\hat{\bkappa} = (\hat{\kappa}_1, \cdots, \hat{\kappa}_p)^{T}$ satisfy $\max_j|\hat{\kappa}_j| \leq 1$, and $\hat{\kappa}_j = \operatorname{sign}(\hat{\beta}_{\lambda,j})$ if $\hat{\beta}_{\lambda,j} \neq 0$. The first-order Taylor expansion of $\bS_n(\hat{\bbeta}_\lambda)$ in the KKT condition at the true value $\bbeta_0$ leads to
$-\dot{\bS}_n(\bbeta_0)(\hat{\bbeta}_\lambda - \bbeta_0) + \lambda\hat{\bkappa} \approx
\bS_n(\bbeta_0).$
It follows that
$\bbeta_\lambda^c - \bbeta_0 \approx \{-\dot{\bS}_n(\bbeta_0) \}^{-1}\bS_n(\bbeta_0),$
where ${\bbeta}^c_\lambda$ is a bias-corrected lasso estimator \citep{van2014asymptotically}:
\begin{equation} \label{eqn:beta_c}  \textstyle
\bbeta_\lambda^c \overset{\text{def}}{=} \hat{\bbeta}_\lambda + \{-\dot{\bS}_n(\bbeta_0) \}^{-1}\lambda \hat{\bkappa} = \hat{\bbeta}_\lambda + \{-\dot{\bS}_n(\bbeta_0) \}^{-1}\bS_n(\hat\bbeta_\lambda).
\end{equation}
The second equality in \eqref{eqn:beta_c} follows directly from the KKT condition and the definition of the sensitivity matrix $-\dot{\bS}_n(\bbeta) = \frac{1}{n\phi}\sum_{i=1}^{n}  v(\mu_i) \bx_i\bx_i^T$, which is assumed to be a positive-definite Hessian matrix, and $v(\cdot)$ is the variance function.
For now, let us first consider the case when $p < n$.
We show in Theorem~\ref{thm:dist} that under some regularity conditions, $\bbeta^c_\lambda$ is asymptotically normally distributed, namely,
\begin{equation} \label{eqn:aymp_norm}
n^{1/2} ({\bbeta}^c_\lambda -\bbeta_0) \overset{d}{\to} \mathcal{N} (0, \bSigma(\bbeta_0)), \text{ as } n\to \infty,
\end{equation}
where $\bSigma(\bbeta_0) = [ E\{ -\dot{\bS}_n(\bbeta_0)\}]^{-1}$. Based on the joint asymptotic normality in \eqref{eqn:aymp_norm}, following \cite{xie2013confidence}, we form the asymptotic confidence distribution density function of $\bbeta_0$ as
$ {h}_n(\bbeta_0) \propto \exp[ -\frac{n}{2}( \bbeta_0 - {\bbeta}^c_\lambda )^{T}\left\{\bSigma(\bbeta_0)\right\}^{-1} (\bbeta_0 - {\bbeta}^c_\lambda)]$.
Replacing $\bbeta_0$ in \eqref{eqn:beta_c} by the sparse lasso estimator $\hat{\bbeta}_\lambda$, we obtain
\begin{equation} \label{eqn:beta_c_emp}
\hat{\bbeta}^c_\lambda = \hat{\bbeta}_\lambda + \{-\dot{\bS}_n(\hat{\bbeta}_\lambda)\}^{-1}\bS_n(\hat{\bbeta}_\lambda).
\end{equation}
Likewise, replacing $\bbeta_0$ by  $\hat{\bbeta}_\lambda$ in the asymptotic covariance in \eqref{eqn:aymp_norm} leads to a ``data-driven'' asymptotic confidence density
\begin{equation} \label{eqn:emp_cd}  \textstyle
\hat{h}_n(\bbeta_0) \propto \exp\left[-\frac{n}{2}(\bbeta_0 - \hat{\bbeta}^c_\lambda)^{T}\{-\dot{\bS}_n(\hat{\bbeta}_\lambda)\} (\bbeta_0 - \hat{\bbeta}^c_\lambda)\right].
\end{equation}
It is worth pointing out that this bias-corrected estimator in \eqref{eqn:beta_c_emp} is equivalent to a one-step Newton-Raphson updated estimator of the lasso estimator. In the GLM framework, we have
$
\hat{h}_n(\bbeta_0) \propto \exp[-\frac{1}{2\phi} ( \bbeta_0 - \hat{\bbeta}^c_\lambda )^{T}\{\bX^{T}\bP_n(\hat{\bbeta}_\lambda)\bX\}$ $ (\bbeta_0 - \hat{\bbeta}^c_\lambda ) ],
$
where $\bP_n(\bbeta) = \mbox{diag}\left\{v(\mu_1), \dots, v(\mu_n)\right\}$ is the diagonal matrix of the variance functions.
When the dispersion parameter $\phi$ is unknown, e.g., in the linear regression setting, we use a root-$n$ consistent estimator $\hat{\phi} = (n-||\hat{\bbeta}_\lambda||_0)^{-1} \sum_{i=1}^{n} d(y_i,\hat{\mu}_i)$, where $||\bx||_0$ is the number of non-zero entries of vector $\bx$, $\hat{\mu}_i = g^{-1}(\bx_i^{T}\hat{\bbeta})$, and $d(\cdot, \cdot)$ is the unit deviance function; refer to \cite[Chapter 2]{song2007correlated} for details.

\subsection{Examples}

\begin{example} \normalfont Gaussian linear model. Assume $y_i$ follows a normal distribution with mean $\mu_i=\bx_i^T\bbeta$, variance function $v(\mu_i) = 1$, and link function $g(x)=x$. The score function takes the form $\bS_n(\bbeta) = \frac{1}{n}\sum_{i=1}^n\left\{y_i - \bx_i^{T}\bbeta \right\}\bx_i/\phi$. The confidence density function $\hat{h}_n(\bbeta_0)$ in \eqref{eqn:emp_cd} is obtained by plugging in the bias-corrected estimator $\hat{\bbeta}^c_\lambda = \hat{\bbeta}_\lambda + (\bX^{T}\bX)^{-1}\bX^{T}(\bY - \bX\hat{\bbeta}_\lambda)$. Here $ \bP_n(\hat{\bbeta}_\lambda)=\bI_n.$
\end{example}

\begin{example} \normalfont Binomial logistic model. Assume $y_i$ follows a Bernoulli distribution with probability of success $\mu_i \in (0,1)$, variance function $\hat{v}_i = \hat{\mu}_i(1-\hat{\mu}_i)$, link function $g(\mu_i)=\log(\frac{\mu_i}{1-\mu_i})=\bx_i^T\bbeta$ and $\phi=1$. Similarly, we obtain its confidence density $\hat{h}_n(\bbeta_0)$ with $\hat{\bbeta}^c_\lambda = \hat{\bbeta}_\lambda + \{\bX^{T}\bP_n(\hat{\bbeta}_\lambda)\bX\}^{-1}\bX^{T}(\bY - \hat{\bmu} ),$
	where $\hat{\bmu} = (\hat{\mu}_1, \cdots, \hat{\mu}_n)^{T}$, $\hat{\bmu}_i = \exp(\bx_i^T\hat{\bbeta}_\lambda)/\{1+\exp(\bx_i^T\hat{\bbeta}_\lambda)\}$ and $\bP_n(\hat{\bbeta}_\lambda) = \operatorname{diag}(\hat{v}_1, \dots, \hat{v}_n)$.
\end{example}

\begin{example} \normalfont Poisson log-linear model. Assume $y_i$ follows a Poisson distribution with mean $\mu_i$, variance function $v(\mu_i)=\mu_i$, link function $g(\mu_i)=\log(\mu_i)=x_i^T\beta$ and $\phi=1$. We can obtain $\hat{h}_n(\bbeta_0)$ with $\hat{\bbeta}^c_\lambda = \hat{\bbeta}_\lambda + \{\bX^{T}\bP_n(\hat{\bbeta}_\lambda) $ $ \bX\}^{-1}  \bX^{T}\{\bY - \hat{\bmu} \},$
	where $\hat{\bmu} = (\hat{\mu}_1, \cdots, \hat{\mu}_n)^{T}$, $\hat{\mu}_i = \hat{v}_i = \exp(x_i^T\hat{\beta}_\lambda)$ and $\bP_n(\hat{\bbeta}_\lambda) = \operatorname{diag}(\hat{v}_1, \dots, \hat{v}_n)$.
\end{example}

\subsection{Large Sample Property}
From here on, we bring back the subscript $k$ to denote a quantity concerning the $k$th sub-dataset as the results will be carried forward to Section~\ref{sec:dividecombine} where we discuss the combination step. Let $\underline{\sigma}(M)$ and $\overline{\sigma}(M)$ denote the minimum and maximum singular values of a matrix $M$, respectively. Let $c_{\min}$ and $c_{\max}$ be the minimum and maximum across the set of constants $c_{k}$, $k\in \{1,\dots, K\}$. Denote the signal set by $\mathcal{A}_{0, k} = \{j: \beta_{0, k, j} \neq 0\}$ and the non-signal set by $\mathcal{A}^c_{0, k} = \left\{j: \beta_{0, k, j} = 0\right\}$, where $\bbeta_{0, k} = (\beta_{0, k, 1}, \dots, \beta_{0, k, p})^{T}$ is the true coefficient. Here we allow $p \ge n_k$ and $p$ may diverge to infinity. To establish large-sample properties for $\hat{\bbeta}^c_{\lambda_k,k}$ given in \eqref{eqn:beta_c_emp} based on the $k$th sub-dataset $(\bY_k, \bX_k)$, and subsequently the combined estimator in Section~\ref{sec:dividecombine} across all sub-datasets, we postulate the following regularity conditions:

(C1) Assume the score function is unbiased, namely, $ E [\{\bY_k - g^{-1}(\bX_k^{T}\bbeta_{0,k})\} $ $ \bX_k/{\phi_k}] = 0.$

(C2) Assume $0 < b_k \leq \underline{\sigma}(n_k^{-1/2}\bX_k) \leq \overline{\sigma}(n_k^{-1/2}\bX_k) \leq B_k$ for constants $b_k$ and $B_k$, and $\|\bX_k\|_{\infty} \leq D_k$ for some $D_k > 0$, where $\|\bX\|_{\infty} = \max_{i, j}|x_{i,j}|$.

(C3) For some $\psi_{0,k} > 0$, for all $\bbeta$ satisfying $\|\bbeta_{\mathcal{A}^c_{0,k}}\|_1 \leq 3\|\bbeta_{\mathcal{A}_{0,k}}\|_1$, it holds that $\|\bbeta_{\mathcal{A}_{0,k}}\|_1^2 \leq \|\bbeta\|^2_2s_{0,k}/\psi_{0,k}^2$, where $s_{0,k}$ is the number of true signals in $\bbeta_{0, k}$ and $\|\bbeta\|_2^2 = \bbeta^{T}\bbeta$. In addition, assume $\lambda_k = O\{ \sqrt{\log p/n_{k}} \}$ and $s_{0,k} = o\left\{(n_{k}/p)^{1/2}/\log p\right\}$.

(C4) Assume the same underlying true parameters $\bbeta_0 = \bbeta_{0,k}$,  $k \in \{1,\dots, K\}$. Denote the common signal set, non-signal set, and number of signals as $\mathcal{A}_{0} = \mathcal{A}_{0,k}$, $\mathcal{A}_{0}^c = \mathcal{A}_{0,k}^c$ and $s_{0} = s_{0,k}$, respectively, for all $k$. Further, assume $0<b_{\min}<B_{\max}<\infty$ and $\psi_{0, \min} > 0$.

Conditions (C1) and (C2) are two mild regularity conditions widely used in the literature; see for example \cite{liu2015multivariate}. It follows from condition (C2) that, $C \geq \max_{\mu \in \Omega_{\delta}}v(\mu) \geq \min_{\mu \in \Omega_{\delta}}v(\mu) \geq c > 0$ with $\Omega_\delta = \{ g^{-1}(\bx^{T}\bbeta): || \bx^{T}\bbeta - \bx^{T}\bbeta_{0, k} ||_1 < \delta, \bx \in R^{p}\}$ for some positive constants $\delta, c$, and $C$. Condition (C3) is the compatibility condition required to ensure the convergence of lasso estimator in terms of both $\ell_1$ and $\ell_2$ norm \citep{buhlmann2011statistics}. When $p = O(n_k^{\delta})$ with $\delta \in [0, 1)$, condition (C3) states that $s_{0, k}$ must be of the order of $o(n_k^{(1-\delta)/2}/\log p)$ in the GLM, which is slightly stronger than order
$s_0 = o(n_k^{1/2}/\log p)$, a usual condition required in the linear model; see for examples \cite{zhang2014confidence} and \cite{van2014asymptotically} and detailed discussion therein.
Condition (C4) is the model homogeneity assumption as well as the uniformly bounded assumption across $K$ sub-datasets, which is required to combine results, as considered in Theorem~\ref{thm:eff}.

\begin{theorem} \label{thm:dist}
	Under conditions (C1)-(C3), for $p = O(n_k^{\delta})$, $\delta \in [0,1)$, and any fixed integer $q$, let $\bH$ be a matrix of rank $q$ with $\overline{\sigma}(\bH) < \infty$. Then the fixed-length bias-corrected estimator
	$ \hat{\bgamma}_{\lambda_k,k} = \bH\hat{\bbeta}^c_{\lambda_k,k}$, with $\hat{\bbeta}^c_{\lambda_k,k}$ given in \eqref{eqn:beta_c_emp}, is consistent and asymptotically normally distributed, namely,
	$ n_k^{1/2}(\hat{\bgamma}_{\lambda_k,k} - \bgamma_{0,k}) \overset{d}{\to} \mathcal{N}(0, \bJ_{\bgamma, k}(\bbeta_{0, k})),$ as $n_k \to \infty,$
	where $\bgamma_{0,k} = \bH \bbeta_{0,k}$, and $\bJ_{\bgamma, k}(\bbeta_{0,k}) = E\{-\bH $ $ \dot{\bS}^{-1}_{n_k}(\bbeta_{0, k})  \bH^{T}\}$.
\end{theorem}
Theorem~\ref{thm:dist} may be viewed as an extension of the covariate-wise asymptotic result in \cite{van2014asymptotically} to the joint asymptotic distribution on $\hat{\bgamma}_{\lambda_k,k}$, a fixed-length sub-vector of $\hat{\bbeta}^c_{\lambda_k,k}$. Matrix $\bH$ chosen under a target subset of parameters allows to perform a joint inference, and univariate inference is a special case with $q = 1$. We emphasize the need of a joint asymptotic distribution in order to use the method of confidence distribution in \eqref{eqn:emp_cd} to combine results in Section~\ref{sec:dividecombine}.
This is a critical step to yield a combined estimator and related inference.
Corollary~\ref{thm:cor2} establishes the validity of the bias-corrected lasso estimator $\hat{\gamma}_{\lambda_{cv}, k}$ with $\lambda_{cv, k}$ being selected via the commonly used R-fold cross-validation procedure. It shows that such $\lambda_{cv, k}$ satisfies a sufficient condition required by Theorem~\ref{thm:dist}, and can be tuned locally within individual sub-datasets. 	
In effect, when the sample sizes $\{n_k\}_{k=1}^{K}$ are balanced, a single tuning parameter $\lambda$ is needed. However, to synchronously tune a common $\lambda$ across $K$ sub-datasets will introduce additional overhead cost in communication. Thus, we keep parameter tuning separate. More discussion is given in Remark~\ref{remark:mr5} in Section~\ref{sec:dividecombine}.
A brief proof of Theorem~\ref{thm:dist} is given in the Appendix, and the complete proofs of Theorem~\ref{thm:dist} and Corollary~\ref{thm:cor2} are given in the Supplementary Material.

\begin{corollary} \label{thm:cor2}
	Under the same conditions of Theorem~\ref{thm:dist}, for any finite integer $R$, $\hat{\bgamma}_{\lambda_{cv,k}, k}$ has the same asymptotic distribution with the tuning parameter $\lambda_{cv,k}$ being obtained from $R$-fold cross-validation using the $k$th sub-dataset.
\end{corollary}

\begin{remark} \normalfont
	The procedure based on Theorem~\ref{thm:dist} for the construction of the confidence density remains valid when the adaptive lasso estimator \citep{zou2006adaptive} is used to replace $\hat{\bbeta}_\lambda$ in \eqref{eqn:beta_c_emp} and \eqref{eqn:emp_cd}. An adaptive lasso estimator is obtained by
	$\check{\bbeta}_\lambda = {\arg \max}_{\bbeta}\frac{1}{n\phi}\sum_{i=1}^n\{y_{i}\bx_i^T\bbeta - b(\bx_i^T\bbeta)\} - \lambda\sum_{j=1}^p\hat{w}_j|\beta_j|,$
	where the weights $\{\hat{w}_j\}_{j=1}^p$ are given by $\hat{w}_j = (|\hat{\beta}_j^{ini}|)^{-\xi}$, with an initial root-$n$ consistent estimate $\hat{\bbeta}^{ini}$ of $\bbeta$ and some suitable constant $\xi > 0$, which is typically set to 1.
\end{remark}

\begin{remark} \normalfont
	Collinearity is often encountered in high-dimensional data analysis where some of the covariates are highly correlated. One solution is to construct the confidence distribution in \eqref{eqn:emp_cd} by using the KKT condition of the elastic net estimator \citep{zou2005regularization}. Another remedy to improve numerical stability is to use a ridge-type estimator by adding a ridge term $\tau \bI_p$, where $\tau > 0$, to stabilize the matrix inverse of $-\dot{\bS}_n(\bbeta_0)$, i.e., $\{-\dot{\bS}_n(\bbeta_0) + \tau \bI_p\}^{-1}$.
\end{remark}

\section{Combined Estimation and Inference}  \label{sec:dividecombine}
We now consider a full data of size $N$ being partitioned into $K$ sub-datasets, $\{(\bY_k, \bX_k)\}_{k=1}^K$, each with size $n_k$, and $N = \sum_{k=1}^{K}n_k$. Here, both $p$ and $K$ are allowed to diverge along with $N$. Let $n_{\inf} = \inf_{k \in \{1, \dots, K\}}n_k$ be the sample size infimum as $N$ and $K$ grow. Consider a target parameter set $\bgamma = \bH \bbeta$, where $q =\mbox{dim}(\bgamma)$ is fixed. At a rate $p = O(n_{\inf}^{\delta}), \delta \in [0,1)$, we obtain $\hat{\bgamma}_k = \arg \max_{\bgamma = \bH\bbeta} \mathcal{L}_{n_k}(\bbeta; \bY_k, \bX_k)$, where $\mathcal{L}_{n_k}(\bbeta; \bY_k, \bX_k)$ is the log-likelihood function of the $k$th sub-dataset $(\bY_k, \bX_k), k\in \{1,\dots,K\}$.  If there existed a ``god-made'' computer with unlimited computational capacity to store and process the full data, the centralized MLE could be applied directly to obtain
$
\hat{\bgamma}_{mle} =
{\arg\max}_{\bgamma =\bH\bbeta} \mathcal{L}_N (\bbeta; \bY, \bX)=
{\arg\max}_{\bgamma=\bH\bbeta}\sum_{k=1}^K\mathcal{L}_{n_k}(\bbeta; \bY_k, \bX_k),	
$
where $\mathcal{L}_N (\bbeta; \bY, \bX)$ is the log-likelihood function of the full data $(\bY, \bX)$.
Arguably, $\hat{\bgamma}_{mle}$ is the gold standard for inference. There are many ways to combine estimates $\hat{\bgamma}_k$ obtained from sub-datasets.
This paper considers using the confidence distribution due to its generalizability under unified objective functions and its ease  in establishing statistical inferences. For each sub-dataset $(\bY_k, \bX_k)$, we first apply Theorem~\ref{thm:dist} to construct the asymptotic confidence density $\hat{h}_{n_k}(\bgamma_0)$, $k\in\{1,\dots,K\}$. Then, in the same spirit as \cite{liu2015multivariate}, we may combine the $K$ confidence densities to derive a combined estimator of $\bgamma_0$, denoted by $\hat{\bgamma}_{dac}$, where $dac$ refers to divide-and-combine, given as follows:
\begin{equation} \label{eqn:beta_dac_method} \textstyle
\begin{split}
\textstyle \hat{\bgamma}_{dac}
& = \textstyle {\arg\max}_{\bgamma} \log \prod_{k=1}^{K} \hat{h}_{n_k}(\bgamma) \\
& = \textstyle {\arg\min}_{\bgamma} \sum_{k=1}^K \frac{1}{2\hat{\phi}_k}(\bgamma - \hat{\bgamma}_{\lambda_k,k})^{T}
\left[\bH \left\{\bX_k^T\bP_{n_k}(\hat{\bbeta}_{\lambda_k,k})\bX_k\right\}^{-1}  \bH^{T}\right]^{-1}
(\bgamma - \hat{\bgamma}_{\lambda_k,k}),
\end{split}
\end{equation}
where  $\hat{\bgamma}_{\lambda_k, k} = \bH\hat{\bbeta}^c_{\lambda_k, k}$ and $\hat{\bbeta}_{\lambda_k,k}^{c}$ is the estimate given in \eqref{eqn:beta_c_emp} with respect to the $k$th sub-dataset $(\bY_k, \bX_k)$. The key advantage of the approach in \eqref{eqn:beta_dac_method} is to derive an inference procedure for the combined estimator $\hat{\bgamma}_{dac}$, as stated in Theorem~\ref{thm:eff} under diverging $p = O(n_{\inf}^{\delta}), \delta \in [0,1)$.

\begin{theorem} \label{thm:eff}
	Assume $K=O(N^{1/2 - \xi}), \xi \in (0, 1/2]$.  Under conditions (C1)-(C4), if $E\left[\dot{\bS}^{-1}_{n_k}(\bbeta_0)\bS_{n_k}(\bbeta_0)\right] = \bm{0}$ and
	$\bJ_{\gamma}(\bbeta_{0}) \equiv \bJ_{\bgamma, k}(\bbeta_{0, k})$, respectively, for all $k$, then the MODAC estimator $\hat{\bgamma}_{dac}$ obtained from (\ref{eqn:beta_dac_method}) is consistent and asymptotically normally distributed, namely,
	$ N^{1/2}(\hat{\bgamma}_{dac} - \bgamma_0) \overset{d}{\to} \mathcal{N}(0, \bJ_{dac, \bgamma}(\bbeta_0)) $
	as $n_{\inf} \to \infty$, with $\bJ_{dac, \bgamma}(\bbeta_0) = \bJ_{\bgamma}(\bbeta_0)$, where the latter is the centralized Fisher information matrix of the full data. That is, the MODAC $\hat{\bgamma}_{dac}$ is asymptotically as efficient as the centralized MLE $\hat{\bgamma}_{mle}$.
\end{theorem}

The key result of Theorem~\ref{thm:eff} is that the combined estimator $\hat{\bgamma}_{dac}$ and the gold standard MLE $\hat{\bgamma}_{mle}$ are asymptotically equally efficient.  Although it may be tempting to allocate CPUs to speed up computation, the order of $K$ in Theorem \ref{thm:eff} guides us to choose a proper number of CPUs to ensure that each CPU has enough samples. It is worth noting that the dispersion parameter $\phi_k$ is not required to be homogeneous across sub-datasets as it does not affect the estimation; and the divide-and-combine estimator $\hat{\bgamma}_{dac}$ does not require additional conditions than those required by the regularized estimator in each sub-dataset.
This is because constructing confidence densities makes the individual asymptotic normal distributions readily available, and the asymptotic distribution of the combined estimator follows.
The practical implication of Theorem~\ref{thm:eff} is that as long as the sample size of each sub-dataset is not too small, the proposed $\hat{\bgamma}_{dac}$ will have little loss of estimation efficiency, while enjoying fast computing in the analysis of big data. The proof of Theorem~\ref{thm:eff} is given in the Appendix.

For the ease of exposition, without loss of generality, we may take $\bgamma_{dac} = \bbeta_{dac}$, i.e., $q = p$. In this way, we can stick on the notation of $\bbeta$ in the rest of this paper.
By simple algebra, the solution to the divide-and-combine estimator $\hat{\bbeta}_{dac}$ \eqref{eqn:beta_dac_method} can be expressed explicitly as a form of weighted average of $\hat{\bbeta}^c_{\lambda_k,k}, k=1,\dots,K$, as follows:
\begin{equation} \label{eqn:beta_dac}
\textstyle \hat{\bbeta}_{dac} = \{\sum_{k=1}^Kn_k\hat{\bSigma}^{-1}_{n_k}(\hat{\bbeta}_{\lambda_k,k}) \}^{-1}\{\sum_{k=1}^K n_k\hat{\bSigma}^{-1}_{n_k}(\hat{\bbeta}_{\lambda_k,k}) \hat{\bbeta}_{\lambda_k,k}^{c} \},
\end{equation}
where $\hat{\bSigma}^{-1}_{n_k}(\hat{\bbeta}_{\lambda_k,k}) = (n_k \hat{\phi}_k)^{-1} \bX_k^T\bP_{n_k}(\hat{\bbeta}_{\lambda_k,k})\bX_k$. Note that the inverse matrix $\hat{\bSigma}^{-1}_{n_k}(\hat{\bbeta}_{\lambda_k,k})$ in \eqref{eqn:beta_dac} is readily available from the confidence distribution of each sub-dataset. The only matrix inversion required is for the sum of the Fisher information matrices. It follows that the variance-covariance matrix of $\hat{\bbeta}_{dac}$ is estimated by $ \hat{\bSigma}_{dac} = \{\sum_{k=1}^Kn_k\hat{\bSigma}^{-1}_{n_k}(\hat{\bbeta}_{\lambda_k,k}) \}^{-1} $, from which confidence regions for any sub-vector of $\bbeta$ can be obtained by using standard multivariate analysis methods~\citep{johnson2002applied}.

\begin{remark} \label{remard:meta} \normalfont
	Note that when $\lambda_k=0$ for all $k$, our proposed estimator $\hat{\bbeta}_{dac}$ in \eqref{eqn:beta_dac} reduces to the classical meta-estimator $\hat{\bbeta}_{meta} = \{ \sum_{k=1}^{K} n_k\hat{\bSigma}_{n_k}^{-1}(\hat{\bbeta}_k) \}^{-1} $ $ \{\sum_{k=1}^{K} n_k\hat{\bSigma}_{n_k}^{-1}(\hat{\bbeta}_k) \hat{\bbeta}_k \}$. \citet{lin2011aggregated} found a similar result as a special case of the aggregated estimating equation estimator. However, the aggregated estimating equation estimator requires a strong assumption of $K=O(n_{\inf}^{r})$ $(r < 1/3)$, and it does not consider regularized estimation for variable selection. In addition, regardless of $\hat{\bbeta}_{meta}$ and $\hat{\bbeta}_{dac}$ in \eqref{eqn:beta_dac} taking the same form, they are derived from different criteria with different purposes. Specifically, $\hat{\bbeta}_{meta}$ aims to improve statistical power via weighted average, while $\hat{\bbeta}_{dac}$ is obtained by minimizing the combined confidence densities for the interest of statistical inference theory. The flexibility of the confidence density approach allows incorporating additional features in the combination; for example, the homogeneity may be relaxed by imposing a mixture of normals in \eqref{eqn:beta_dac_method}, which is not feasible in the meta-estimator.
\end{remark}

\begin{remark} \label{remard:mv} \normalfont
	A majority voting approach \cite{chen2014split} to combine sparse estimates from $K$ sub-datasets takes the form $
	\hat{\bbeta}_{mv} = \bA\{\sum_{k=1}^{K}n_k\bA^{T}\dot{\bS}_{n_k}(\hat{\bbeta}_k)\bA
	\}^{-1}  \{\sum_{k=1}^{K}n_k\bA^{T}\dot{\bS}_{n_k}(\hat{\bbeta}_k)\bA \hat{\bbeta}_{k, \hat{\mathcal{\bA}}^{(v)}}
	\},$
	where  $\hat{\mathcal{A}}^{(v)} = \{j: \sum_{k=1}^{K}I(\hat{\beta}_{k,j} \neq 0) > w\}$ is a set of selected signals in terms of a prespecified voting threshold $w \in [0,K)$, $\hat{\bbeta}_{k, \hat{\mathcal{A}}^{(v)}}$ denotes a corresponding sub-vector of the lasso estimate $\hat{\bbeta}_k$, and $\bA$ is a $p\times |\hat{\mathcal{A}}^{(v)}|$ subsetting matrix corresponding to set $\hat{\mathcal{A}}^{(v)}$.	The majority voting estimator $\hat{\bbeta}_{mv}$ has been shown to have the oracle property, which, however, is not applicable to statistical inference.
\end{remark}

\begin{remark} \label{remark:mr5} \normalfont
	The role of tuning in individual datasets is not to induce sparsity in the final aggregated estimate, but to produce intermediate sparse estimates that give rise to a robust approximation of the covariance in the individual confidence distributions. Since the bias-correction procedure offsets the effect of sparsity tuning, the choice of tuning parameter becomes of little relevance to the means of the derived confidence distributions. The purpose of our integrative inference distinguishes from those estimation methods given in \cite{lee2017communication,wang2019fast} that aim to produce aggregated sparse estimates, in which a common tuning parameter has to be chosen across all $K$ sub-datasets. As a result, their estimation methods require one more round of synchronization, whereas in ours, tuning can be done in parallel (from Corollary~\ref{thm:cor2}).
\end{remark}

The overall computational complexity of centralized MLE based on Fisher's scoring is of order $O(Np^{2+\epsilon}), \epsilon \in (0,1)$ \citep{toulis2015scalable}, which is dominated by the cost of matrix inversion.
The complexity of divided procedures in MODAC involves coordinate descent (of order $O(2np)$ when $\lambda$ is given \citep{friedman2010regularization}) and evaluating Fisher information matrix (of order $O(np^2)$), for each sub-dataset. The aggregation step involves summation of order $O(Kp^2)$ and matrix inversion of order $O(p^{2 + \epsilon})$. Therefore, the complexity under the ideal parallel situation is of order $O(2np + np^2 + Kp^2 + p^{2 + \epsilon})$. Even in the worst scenario when parallel procedures are run sequentially, the upper bound of overall complexity of MODAC is $O(Knp^2)$, which remains comparable to that of the centralized MLE.
Similarly, the complexity of the distributed meta-estimator is at order $O(np^{2+\epsilon} + Kp^2 + p^{2+\epsilon})$ with an upper bound $O(Knp^{2+\epsilon})$.
The value $\epsilon$ is purely dependent on the choice of a matrix inversion algorithm, and it ranges over $(0.3, 0.4)$ for some efficient algorithms.

\section{Simulation Studies} \label{sec:simulation}
In this section, we demonstrate the numerical performance of our method under linear, logistic and Poisson regressions through simulation experiments. Specifically, we compare across three divide-and-combine methods, including the meta-analysis method by inverse variance weighted averaging described in Remark~\ref{remard:meta}, the majority voting method described in Remark~\ref{remard:mv}, and our method. Note that when $K=1$, under no data partition, meta-analysis is equivalent to the centralized MLE, the majority voting method is equivalent to the centralized lasso regression \citep{tibshirani1996regression}, and our method is equivalent to centralized lasso with post-selection inference from Theorem~\ref{thm:dist}.

All methods are compared thoroughly on the performance of variable selection, statistical inference and computation time. The evaluation metrics for variable selection include the sensitivity and specificity of correctly identifying non-zero coefficients. The evaluation metrics for statistical inference include mean squared error, absolute bias, coverage probability and asymptotic standard error of coefficients in the signal set $\mathcal{A}_0$ and the non-signal set $\mathcal{A}_0^c$, respectively. Coverage probabilities and standard errors are not reported for the majority voting method since it does not provide inference.
We use results from the centralized MLE, $\hat{\bbeta}_{mle}$, as our gold standard in all comparisons. In order to ensure the best variable selection results of the majority voting method, we carefully select $\omega$ in $\hat{\bbeta}_{mv}$ such that the sum of sensitivity and specificity is maximized.
The computation time of all methods includes the time taken to read data from disks to memory and the time taken by numerical calculations. Under the divide-and-combine setting with $K>1$, computation time is reported as the sum of the maximum time used among parallelized jobs and the time used to combine results.
{\color{black}
	All shrinkage estimates are obtained by applying the R package \verb|glmnet| with tuning parameter $\lambda_k$ selected to yield the smallest average 10-fold cross-validated error.}
All simulation experiments are conducted by R software on a standard Linux cluster with 16 GB of random-access memory per CPU.

Table~\ref{tab:simulation1} presents the simulation results from a moderate size dataset with $N=50,000$ and $p=300$ so that methods without data partition can be repeated in multiple rounds of simulations within a reasonable amount of time. Clearly, this is a typical regression data setting with $N \gg p$. We consider linear, logistic and Poisson models, with responses generated from the mean model $E(y_i) = g^{-1}(\bx_{i}^T \bbeta_0), \ i\in\{1,\dots,N\}$, with covariates $\{\bx_i\}_{i=1}^{N}$ generated from the multivariate normal distribution with mean zero and variance one, and with a compound symmetric covariance structure with correlation $\rho=0.8$, a simulation setting similar to that considered by \cite{van2014asymptotically}.  We report scenarios when the full dataset is randomly divided into $K=25$ and $K = 100$ subsets of equal sizes, each with sample size $n_k=2,000$ and $n_k=500$, respectively. Results for $K=1$ are also reported. We randomly select $s_0=10$ coefficients from $\bbeta_0$ to be set at non-zero. The non-zero coefficients are set to $0.3$ for linear models, $0.3$ for logistic models, and $0.1$ for Poisson models. CMLE, META, MV and MODAC denote the centralized MLE, meta-analysis, majority voting and our method of divide-and-combine (MODAC), respectively. Results are based on 500 replications.

\begin{figure}
	\centering
	\begin{subfigure}[b]{0.32\textwidth}
		\centering
		\caption{Linear}
		\label{fig:mseSa}
		\includegraphics[width=\textwidth]{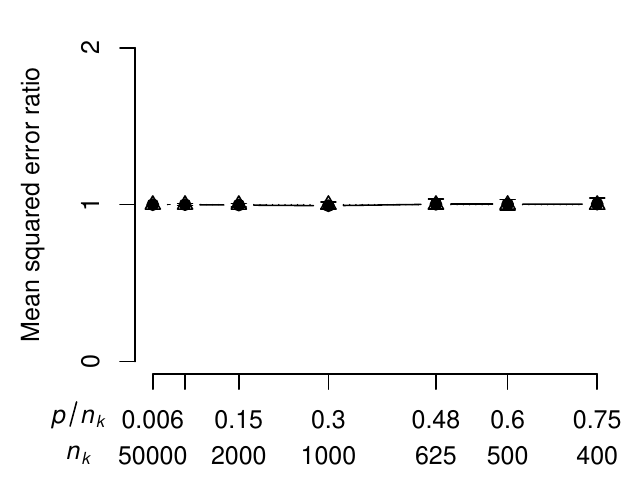}
	\end{subfigure}
	\begin{subfigure}[b]{0.32\textwidth}
		\centering
		\caption{Logistic}	
		\label{fig:mseSb}		
		\includegraphics[width=\textwidth]{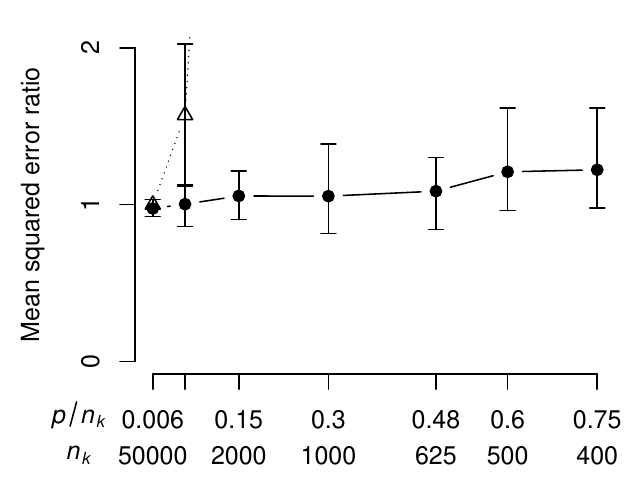}
	\end{subfigure}
	\begin{subfigure}[b]{0.32\textwidth}
		\centering
		\caption{Poisson}	
		\label{fig:mseSc}		
		\includegraphics[width=\textwidth]{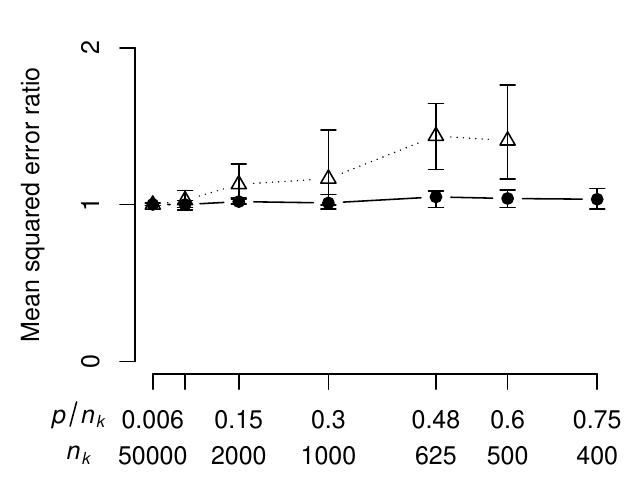}
	\end{subfigure}
	\caption{The $y$-axis measures the ratio of mean squared error over that of the gold standard CMLE, for regression coefficients in set $\mathcal{A}_0$. Median and interquartile ranges of the ratios of META (triangles) and MODAC (solid dots) are shown as the ratio $p/n_k$ increases. We fix $N$ at $50,000$ and $p$ at $300$. META fails to converge for logistic and Poisson regressions when $p/n_k$ is large and the results are unavailable.}
	\label{fig:mseratioS}
\end{figure}

\begin{figure}
	\centering
	\begin{subfigure}[b]{0.32\textwidth}
		\centering
		\caption{Linear}
		\label{fig:covSa}
		\includegraphics[width=\textwidth]{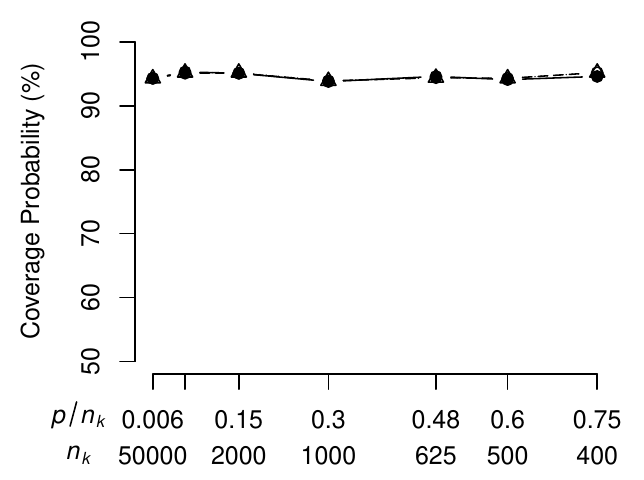}
	\end{subfigure}
	\begin{subfigure}[b]{0.32\textwidth}
		\centering
		\caption{Logistic}	
		\label{fig:covSb}		
		\includegraphics[width=\textwidth]{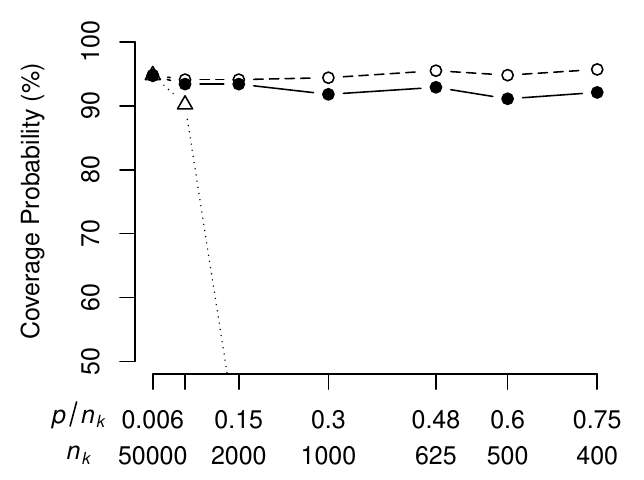}
	\end{subfigure}
	\begin{subfigure}[b]{0.32\textwidth}
		\centering
		\caption{Poisson}	
		\label{fig:covSc}		
		\includegraphics[width=\textwidth]{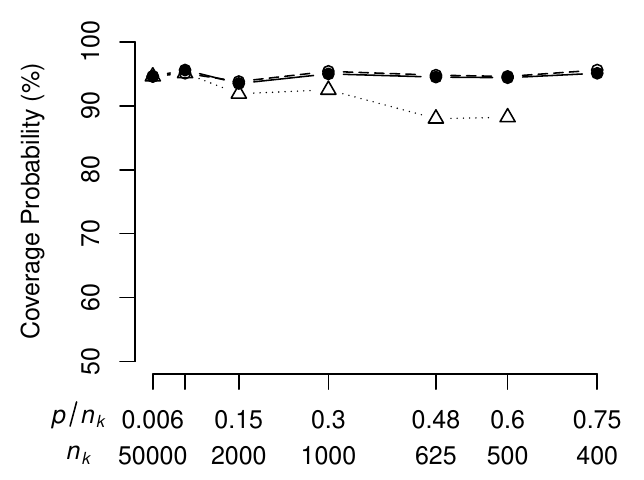}
	\end{subfigure}
	\caption{Coverage probabilities of regression coefficients in set $\mathcal{A}_0$ for the gold standard CMLE (open dots), META (triangles) and MODAC (solid dots) as the ratio of $p$ and $n_k$ increases. The total sample size $N$ and number of covariates $p$ are fixed at $50,000$ and $300$, respectively, for all cases. META fails to converge for logistic regression when $p/n_k \ge 0.3$ and the results are unavailable.}
	\label{fig:covS}
\end{figure}

The results of the Gaussian linear model in Table~\ref{tab:simulation1} reassuringly show that all methods perform as well as the gold standard. META and MODAC exhibit identical performances as that of CMLE regardless of the choices of $K$. This is because under the linear model, CMLE can be directly parallelized, so META and MODAC solutions are exact and identical to CMLE. Among all methods, MV has the highest sensitivity and specificity when $\omega=12$ for $K=25$ and $\omega=50$ for $K=100$. This shows the improvement of selection consistency by divide-and-combine. Under the same model settings, Figs.~\ref{fig:mseratioS} and \ref{fig:covS} display additional simulation results at varying choices of $K$ with $N$ fixed at $50,000$, summarized over 100 replications. Fig.~\ref{fig:mseratioS} shows the ratio comparison of mean squared error of META and MODAC, respectively, to that of CMLE, for $\hat{\bbeta}_{\mathcal{A}_0}$ as $K$ increases. Fig.~\ref{fig:covS} compares the coverage probabilities of $\bbeta_{\mathcal{A}_0}$ between CMLE, META and MODAC. Since META and MODAC are identical to CMLE, their mean-squared errors and coverage probabilities are almost identical, as shown in Figs.~\ref{fig:mseSa} and \ref{fig:covSa}.

\begin{table}
	\centering
	\caption{Simulation results, summarized from 500 replications, under the setting of $N=50,000$ and $p=300$ for linear, logistic and Poisson models. Methods with different $K$ are compared. CMLE denotes the centralized MLE method; META denotes the meta-analysis method; MV denotes the majority voting method; and MODAC denotes the proposed method of divide-and-combine.}
	\label{tab:simulation1}	
	\resizebox{1.0\columnwidth}{!}{
		\begin{tabular}{lrrrrrrrrr}
			\hline
			\multicolumn{10}{c}{Linear Model} \\
			& CMLE & META & META & MV & MV & MV & MODAC & MODAC & MODAC \\ 		
			& ($K=1$) & ($K=25$) & ($K=100$) & ($K=1$) & ($K=25$) & ($K=100$) & ($K=1$) & ($K=25$) & ($K=100$) \\ 		
			&  &  &  &  & ($\omega=12$) & ($\omega=50$) &  &  &  \\
			\hline
			Sensitivity & 1.00 & 1.00 & 1.00 & 1.00 & 1.00 & 1.00 & 1.00 & 1.00 & 1.00 \\
			Specificity & 0.95 & 0.95 & 0.95 & 0.91 & 1.00 & 1.00 & 0.95 & 0.95 & 0.95 \\
			MSE of $\hat{\bbeta}_{\mathcal{A}_0}$ ($\times 100$) & 0.01 & 0.01 & 0.01 & 0.03 & 0.01 & 0.01 & 0.01 & 0.01 & 0.01 \\
			MSE of $\hat{\bbeta}_{\mathcal{A}_0^c}$ ($\times 100$) & 0.01 & 0.01 & 0.01 & 0.00 & 0.00 & 0.00 & 0.01 & 0.01 & 0.01 \\
			Absolute bias of $\hat{\bbeta}_{\mathcal{A}_0}$ & 0.01 & 0.01 & 0.01 & 0.01 & 0.01 & 0.01 & 0.01 & 0.01 & 0.01 \\
			Absolute bias of $\hat{\bbeta}_{\mathcal{A}_0^c}$ & 0.01 & 0.01 & 0.01 & 0.00 & 0.00 & 0.00 & 0.01 & 0.01 & 0.01 \\
			Cov. prob. of $\bbeta_{\mathcal{A}_0}$ & 0.95 & 0.95 & 0.95 & --- & --- & --- & 0.95 & 0.95 & 0.95 \\
			Cov. prob. of $\bbeta_{\mathcal{A}_0^c}$ & 0.95 & 0.95 & 0.95 & --- & --- & --- & 0.95 & 0.95 & 0.95 \\
			Asymp. st. err. of $\hat{\bbeta}_{\mathcal{A}_0}$ & 0.01 & 0.01 & 0.01 & --- & --- & --- & 0.01 & 0.01 & 0.01 \\
			Asymp. st. err. of $\hat{\bbeta}_{\mathcal{A}_0^c}$ & 0.01 & 0.01 & 0.01 & --- & --- & --- & 0.01 & 0.01 & 0.01 \\
			Computation time & 34.85 & 0.62 & 0.20 & 31.50 & 2.16 & 2.08 & 36.61 & 2.28 & 2.14 \\ [6pt]
			\hline
			\multicolumn{10}{c}{Logistic Model} \\
			& CMLE & META & META & MV & MV & MV & MODAC & MODAC & MODAC \\ 		
			& ($K=1$) & ($K=25$) & ($K=100$) & ($K=1$) & ($K=25$) & ($K=100$) & ($K=1$) & ($K=25$) & ($K=100$) \\ 	
			&  &  &  &  & ($\omega=7$) & ($\omega=20$) &  &  &  \\ 	
			\hline
			Sensitivity & 1.00 & 1.00 & 0.00 & 1.00 & 1.00 & 1.00 & 1.00 & 1.00 & 1.00 \\
			Specificity & 0.95 & 1.00 & 1.00 & 0.89 & 1.00 & 1.00 & 0.95 & 0.95 & 0.96 \\
			MSE of $\hat{\bbeta}_{\mathcal{A}_0}$ ($\times 100$) & 0.08 & 0.57 & 189.38 & 0.23 & 0.20 & 0.29 & 0.08 & 0.09 & 0.10 \\
			MSE of $\hat{\bbeta}_{\mathcal{A}_0^c}$ ($\times 100$) & 0.08 & 0.05 & 4.15 & 0.00 & 0.00 & 0.00 & 0.08 & 0.08 & 0.07 \\
			Absolute bias of $\hat{\bbeta}_{\mathcal{A}_0}$ & 0.02 & 0.07 & 1.36 & 0.04 & 0.04 & 0.05 & 0.02 & 0.02 & 0.02 \\
			Absolute bias of $\hat{\bbeta}_{\mathcal{A}_0^c}$ & 0.02 & 0.02 & 0.16 & 0.00 & 0.00 & 0.00 & 0.02 & 0.02 & 0.02 \\
			Cov. prob. of $\bbeta_{\mathcal{A}_0}$ & 0.95 & 0.36 & 1.00 & --- & --- & --- & 0.95 & 0.94 & 0.92 \\
			Cov. prob. of $\bbeta_{\mathcal{A}_0^c}$ & 0.95 & 1.00 & 1.00 & --- & --- & --- & 0.95 & 0.95 & 0.96 \\
			Asymp. st. err. of $\hat{\bbeta}_{\mathcal{A}_0}$ & 0.03 & 0.03 & 1895.12 & --- & --- & --- & 0.03 & 0.03 & 0.03 \\
			Asymp. st. err. of $\hat{\bbeta}_{\mathcal{A}_0^c}$ & 0.03 & 0.03 & 1893.23 & --- & --- & --- & 0.03 & 0.03 & 0.03 \\
			Computation time & 66.01 & 1.63 & 1.40 & 260.48 & 15.78 & 10.42 & 266.09 & 15.92 & 10.53 \\ [6pt]
			\hline
			\multicolumn{10}{c}{Poisson Model} \\
			& CMLE & META & META & MV & MV & MV & MODAC & MODAC & MODAC \\ 		
			& ($K=1$) & ($K=25$) & ($K=100$) & ($K=1$) & ($K=25$) & ($K=100$) & ($K=1$) & ($K=25$) & ($K=100$) \\ 	
			&  &  &  &  & ($\omega=7$) & ($\omega=26$) &  &  &  \\ 	
			\hline
			Sensitivity & 1.00 & 1.00 & 1.00 & 1.00 & 1.00 & 1.00 & 1.00 & 1.00 & 1.00 \\
			Specificity & 0.95 & 0.94 & 0.91 & 0.91 & 1.00 & 1.00 & 0.95 & 0.95 & 0.95 \\
			MSE of $\hat{\bbeta}_{\mathcal{A}_0}$ ($\times 100$) & 0.70 & 0.80 & 0.90 & 1.70 & 0.80 & 0.50 & 0.70 & 0.70 & 0.70 \\
			MSE of $\hat{\bbeta}_{\mathcal{A}_0^c}$ ($\times 100$) & 0.70 & 0.70 & 0.80 & 0.00 & 0.10 & 0.00 & 0.70 & 0.70 & 0.70 \\
			Absolute bias of $\hat{\bbeta}_{\mathcal{A}_0}$ & 0.01 & 0.01 & 0.01 & 0.01 & 0.01 & 0.00 & 0.01 & 0.01 & 0.01 \\
			Absolute bias of $\hat{\bbeta}_{\mathcal{A}_0^c}$ & 0.01 & 0.01 & 0.01 & 0.00 & 0.00 & 0.00 & 0.01 & 0.01 & 0.01 \\
			Cov. prob. of $\bbeta_{\mathcal{A}_0}$ & 0.95 & 0.93 & 0.90 & --- & --- & --- & 0.95 & 0.95 & 0.95 \\
			Cov. prob. of $\bbeta_{\mathcal{A}_0^c}$ & 0.95 & 0.94 & 0.91 & --- & --- & --- & 0.95 & 0.95 & 0.95 \\
			Asymp. st. err. of $\hat{\bbeta}_{\mathcal{A}_0}$ & 0.01 & 0.01 & 0.01 & --- & --- & --- & 0.01 & 0.01 & 0.01 \\
			Asymp. st. err. of $\hat{\bbeta}_{\mathcal{A}_0^c}$ & 0.01 & 0.01 & 0.01 & --- & --- & --- & 0.01 & 0.01 & 0.01 \\
			Computation time & 42.26 & 1.46 & 0.40 & 132.06 & 26.57 & 25.00 & 136.85 & 26.67 & 25.08 \\ 	
			\hline
		\end{tabular}
	}
\end{table}

The existence of exact solutions for divide-and-combine methods under the linear model no longer holds in other generalized linear models, where iterative numerical procedures are needed to search for the estimates. For example in the logistic model, the $p/n_k$ ratio is responsible for numerical stability, as shown in Figs.~\ref{fig:mseSb} and \ref{fig:covSb}. When $p/n_k$ approaches one, the mean squared errors and coverage probabilities of  META quickly deviate from those of  CMLE, whereas MODAC remains stable. Although $p$ is much smaller than $N$, data partitioning may sometimes result in $p$ closer to $n_k$ for some sub-datasets.
Regularization is shown in our simulation to be an appealing strategy to reduce the dimension of the optimization to achieve more stable numerical performance.
The regularization helps stabilize the Newton-Raphson iterative updating algorithm, in which the Hessian matrix may be otherwise poorly estimated in case of $p/n_k$ being close to one.
\textcolor{black}{The numerical results of META appear to be unstable within each sub-dataset in both cases $K=25$ and $K=100$. Such poor numerical performance results from the estimated probabilities $\hat{\mu}_i$ approaching the boundaries in $[0,1]$, causing the variance estimates $\hat{\mu}_i(1-\hat{\mu}_i)$ too close to 0. In short, META gives biased parameter estimates and overestimated standard errors of these parameter estimates, and is very sensitive to the choice of $K$.} On the other hand, through the regularized estimation of $\mu_i$, the proposed MODAC exhibits stable performance similar to that of CMLE. The bias of MV for $\mathcal{A}_0^c$ is higher than that of CMLE as expected due to the $\ell_1$ penalty.

In regard to the Poisson model, Table~\ref{tab:simulation1} shows that similar to our findings in the linear and logistic models, MODAC again gives the most stable results among all divide-and-combine methods. On the other hand, META gives improper coverage probabilities in comparison to the nominal 95\% level, as well as poorer selection accuracy than CMLE. In Fig.~\ref{fig:mseSc}, we see that the mean squared errors of MODAC is stable against the change of $p/n_k$. In contrast, the mean squared errors of META quickly deviates from the mean squared error of CMLE for the Poisson model as $p/n_k$ increases. In Fig.~\ref{fig:covSc}, as similar to the logistic model, the 95\% confidence interval coverage probabilities of MODAC remains close to the nominal level, whereas the coverage probabilities of META deviates from 95\% when $p/n_k$ gets close to one.
MV gives the best variable selection with $\omega$ carefully chosen.

\begin{figure}
	\centering
	\begin{subfigure}[b]{0.32\textwidth}
		\centering
		\caption{Linear}
		\label{fig:comptimea}		
		\includegraphics[width=\textwidth]{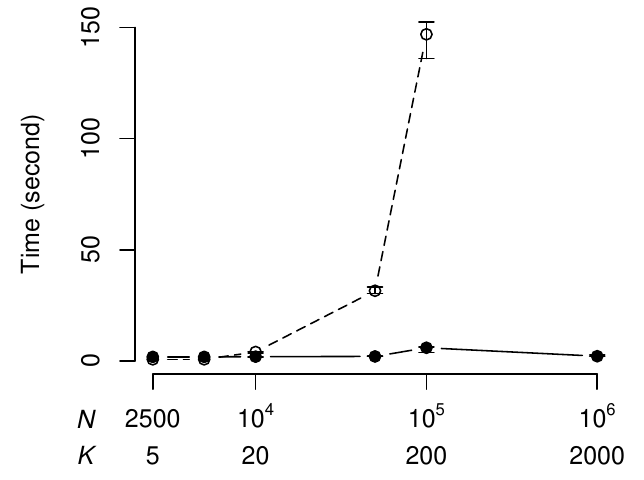}
	\end{subfigure}
	\begin{subfigure}[b]{0.32\textwidth}
		\centering
		\caption{Logistic}	
		\label{fig:comptimeb}		
		\includegraphics[width=\textwidth]{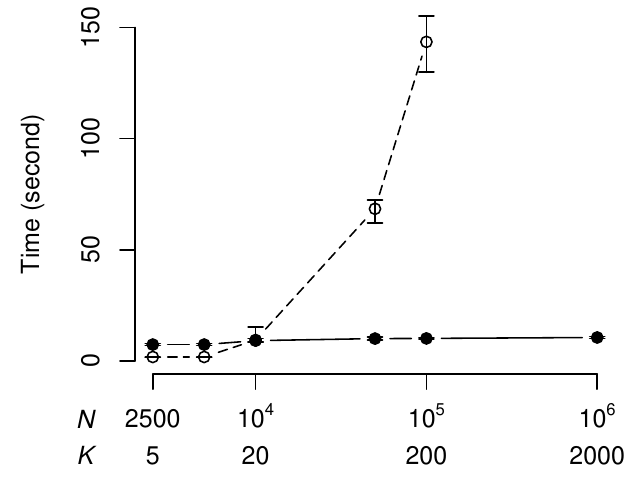}
	\end{subfigure}
	\begin{subfigure}[b]{0.32\textwidth}
		\centering
		\caption{Poisson}	
		\label{fig:comptimec}		
		\includegraphics[width=\textwidth]{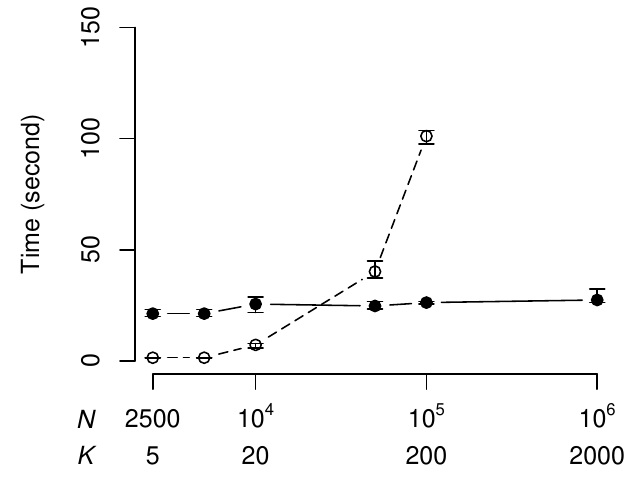}
	\end{subfigure}
	\caption{Median computation time and interquartile ranges for CMLE (open dots) and MODAC (solid dots) as $N$ increases. The sample size of each sub-dataset $n_k$ in our method is fixed at $500$ by increasing $K$. CMLE fails when $N=10^6$ due to memory limitation and the results are unavailable.}
	\label{fig:comptime}
\end{figure}

The key message from Table~\ref{tab:simulation1} and Figs. \ref{fig:mseratioS}-\ref{fig:covS} is that the invocation of regularization greatly helps to achieve consistent and stable mean and variance estimation in the application of divide-and-combine methods. We see that MODAC produces the most comparable results to those of the gold standard, and is virtually unaffected by the partition size $K$. In contrast, the performances of META and MV vary over the partition size $K$. Another noticeable advantage of MODAC is the saving of computation time in comparison to gold standard CMLE due to MODAC's scalability, as shown in Fig. \ref{fig:comptime} with increase in $N$ and $n_k=500$ in MODAC, based on 100 replications.  We see that the computational burden increases sharply for CMLE as $N$ increases, whereas the computation time for MODAC remains almost the same in all three types of models, which clearly demonstrates its scalability. Computation time for CMLE when $N = 10^6$ is not reported because the computation exceeds the maximum memory limit allowed on the Linux cluster.  In summary, MODAC achieves significant computation time reduction without sacrificing statistical accuracy.  Despite the fact that META is the fastest as it does not involve a tuning parameter selection step, its results are clearly unstable in both the logistic and Poisson models.
We present additional simulations in the Supplementary Material to show (i) sensitivity of MV regarding choices of $\omega$, (ii) sensitivity of MODAC under different levels of correlation in design matrices, and (iii) comparison with another faster version of CMLE given by R package \verb|speedglm|.

\section{Real Data Application} \label{sec:Application}

\begin{table}
	\centering
	\caption{Estimation and inference results of association study between potential risk factors and binary injury outcome. Logistic model is fitted using the centralized maximum likelihood estimation method (CMLE), the meta-analysis method (META), and our proposed method of divide-and-combine (MODAC). Run time is presented in square brackets.}
	\label{tab:application}	
	\resizebox{1.0\columnwidth}{!}{
		\begin{tabular}{lrrrrrrrrr}
			\hline
			& \multicolumn{3}{c}{CMLE (1.17s)} & \multicolumn{3}{c}{META (0.03s)} & \multicolumn{3}{c}{MODAC (0.62s)} \\
			Predictors & Estimate & St. Err. & $p$-value & Estimate & St. Err. & $p$-value & Estimate & St. Err. & $p$-value \\
			\hline
			Age & 0.08 & 0.01 & 0.00 & 0.08 & 0.01 & 0.00 & 0.08 & 0.01 & 0.00 \\
			If any other passenger & -0.17 & 0.03 & 0.00 & -0.16 & 0.03 & 0.00 & -0.17 & 0.03 & 0.00 \\
			If passenger below 14 & -0.31 & 0.06 & 0.00 & -0.27 & 0.06 & 0.00 & -0.26 & 0.05 & 0.00 \\
			If driver female & -0.08 & 0.03 & 0.01 & -0.08 & 0.03 & 0.02 & -0.08 & 0.03 & 0.01 \\
			Driver weight & 0.10 & 0.01 & 0.00 & 0.09 & 0.01 & 0.00 & 0.10 & 0.01 & 0.00 \\
			Driver height & -0.09 & 0.02 & 0.00 & -0.08 & 0.02 & 0.00 & -0.09 & 0.02 & 0.00 \\
			If restraint used & -1.07 & 0.03 & 0.00 & -1.00 & 0.03 & 0.00 & -1.05 & 0.03 & 0.00 \\
			Number of lanes & 0.03 & 0.01 & 0.03 & 0.03 & 0.01 & 0.04 & 0.03 & 0.01 & 0.03 \\
			Speed limit & 0.01 & 0.01 & 0.65 & 0.00 & 0.01 & 0.81 & 0.00 & 0.01 & 0.72 \\
			Vehicle age & 0.01 & 0.01 & 0.43 & 0.01 & 0.01 & 0.40 & 0.01 & 0.01 & 0.40 \\
			Vehicle curb weight & -0.02 & 0.02 & 0.30 & -0.01 & 0.02 & 0.48 & -0.02 & 0.02 & 0.34 \\ 				
			If truck & -0.05 & 0.04 & 0.19 & -0.05 & 0.04 & 0.21 & -0.05 & 0.04 & 0.18 \\ 				
			If vehicle in previous accident & -0.11 & 0.03 & 0.00 & -0.10 & 0.03 & 0.00 & -0.10 & 0.03 & 0.00 \\
			If four wheel drive & 0.01 & 0.04 & 0.69 & 0.02 & 0.04 & 0.67 & 0.01 & 0.04 & 0.70 \\
			If drinking involved & 0.00 & 0.04 & 0.90 & 0.01 & 0.05 & 0.78 & 0.00 & 0.04 & 0.90  \\
			If drug involved & 0.03 & 0.04 & 0.51 & 0.03 & 0.05 & 0.49 & 0.03 & 0.04 & 0.54 \\ 		
			If Hispanic & 0.12 & 0.04 & 0.00 & 0.11 & 0.04 & 0.00 & 0.11 & 0.04 & 0.00 \\
			If roadway condition bad & 0.00 & 0.05 & 0.98 & 0.03 & 0.05 & 0.60 & 0.00 & 0.05 & 0.98 \\
			If inclement weather & -0.02 & 0.06 & 0.77 & -0.03 & 0.06 & 0.58 & -0.02 & 0.06 & 0.77 \\ 				
			Driver race - White (baseline) \\
			\rowcolor{Gray}
			Driver race - Black & -0.07 & 0.03 & 0.03 & -0.06 & 0.03 & 0.07 & -0.07 & 0.03 & 0.03 \\
			Driver race - Asian & -0.08 & 0.07 & 0.23 & -0.01 & 0.07 & 0.83 & -0.08 & 0.07 & 0.23 \\
			Region - West (baseline) \\
			Region - Mid-Atlantic & -0.16 & 0.04 & 0.00 & -0.15 & 0.04 & 0.00 & -0.15 & 0.04 & 0.00 \\
			Region - Northeast & -0.07 & 0.06 & 0.22 & -0.04 & 0.06 & 0.52 & -0.07 & 0.06 & 0.23 \\
			Region - Northwest & 0.27 & 0.05 & 0.00 & 0.26 & 0.05 & 0.00 & 0.28 & 0.05 & 0.00 \\
			Region - South & -0.29 & 0.05 & 0.00 & -0.26 & 0.05 & 0.00 & -0.27 & 0.04 & 0.00 \\
			Region - Southeast & -0.29 & 0.06 & 0.00 & -0.25 & 0.06 & 0.00 & -0.26 & 0.06 & 0.00 \\
			Region - Southwest & -0.13 & 0.04 & 0.00 & -0.12 & 0.04 & 0.00 & -0.12 & 0.04 & 0.00 \\
			Light condition - daylight (baseline) \\
			Light condition - dark & 0.05 & 0.05 & 0.24 & 0.07 & 0.05 & 0.16 & 0.05 & 0.04 & 0.26 \\
			Light condition - dawn/dusk & -0.02 & 0.06 & 0.76 & 0.03 & 0.07 & 0.71 & -0.02 & 0.06 & 0.76 \\
			Light condition - dark/lighted & -0.03 & 0.03 & 0.33 & -0.02 & 0.03 & 0.45 & -0.03 & 0.03 & 0.33 \\
			Season - Summer (baseline) \\
			Season - Spring & 0.12 & 0.03 & 0.00 & 0.11 & 0.04 & 0.00 & 0.11 & 0.03 & 0.00 \\ 				
			Season - Fall  & 0.01 & 0.04 & 0.83 & 0.00 & 0.04 & 0.93 & 0.00 & 0.03 & 0.88 \\
			Season - Winter & 0.03 & 0.04 & 0.34 & 0.03 & 0.04 & 0.46 & 0.03 & 0.04 & 0.37 \\
			Trafficway flow - divided with barrier (baseline) \\
			Trafficway flow - divide without barrier & 0.02 & 0.04 & 0.64 & 0.01 & 0.04 & 0.73 & 0.01 & 0.04 & 0.71 \\
			Trafficway flow - not divided  & -0.02 & 0.04 & 0.63 & -0.03 & 0.04 & 0.49 & -0.02 & 0.04 & 0.53 \\
			Trafficway flow - one way & -0.19 & 0.06 & 0.00 & -0.16 & 0.06 & 0.01 & -0.17 & 0.06 & 0.00 \\
			Day of Week - Sun (baseline)\\
			Day of week - Mon & -0.21 & 0.05 & 0.00 & -0.19 & 0.05 & 0.00 & -0.21 & 0.04 & 0.00 \\
			Day of week - Tue & -0.22 & 0.05 & 0.00 & -0.21 & 0.05 & 0.00 & -0.21 & 0.04 & 0.00 \\
			\rowcolor{Gray}
			Day of week - Wed & -0.09 & 0.04 & 0.03 & -0.09 & 0.05 & 0.06 & -0.09 & 0.04 & 0.03 \\
			Day of week - Thu & -0.17 & 0.04 & 0.00 & -0.17 & 0.05 & 0.00 & -0.17 & 0.04 & 0.00 \\
			Day of week - Fri & -0.15 & 0.04 & 0.00 & -0.15 & 0.04 & 0.00 & -0.15 & 0.04 & 0.00 \\ 				
			Day of week - Sat & -0.19 & 0.04 & 0.00 & -0.18 & 0.04 & 0.00 & -0.18 & 0.04 & 0.00 \\
			Year - 2009 (baseline) \\
			Year - 2010 & -0.06 & 0.04 & 0.15 & -0.05 & 0.04 & 0.26 & -0.05 & 0.04 & 0.18 \\
			Year - 2011 & 0.01 & 0.04 & 0.78 & 0.01 & 0.04 & 0.83 & 0.01 & 0.04 & 0.81 \\
			Year - 2012 & 0.11 & 0.04 & 0.01 & 0.11 & 0.04 & 0.01 & 0.10 & 0.04 & 0.01 \\
			Year - 2013 & 0.08 & 0.04 & 0.08 & 0.07 & 0.04 & 0.09 & 0.07 & 0.04 & 0.09 \\
			Year - 2014 & 0.04 & 0.05 & 0.32 & 0.06 & 0.05 & 0.22 & 0.04 & 0.04 & 0.34 \\
			Year - 2015 & 0.14 & 0.05 & 0.00 & 0.15 & 0.05 & 0.00 & 0.14 & 0.05 & 0.00 \\
			\hline
		\end{tabular}
	}
\end{table}

We illustrate our method using a publicly available dataset from the National Highway and National Automotive Sampling System Crashworthiness Data System between the years of 2009 and 2015. Details on the access of the data are provided in the Supplementary Material. This national database contains detailed information of about 5,000 crashes each year sampled across the United States. The response variable of interest is a binary outcome of injury severity, where 1 corresponds to a crash leading to moderate or severer injury, and 0 for minor or no injury. Most of the predictors included in this study are categorical, and are transformed into dummy variables before regression. Our logistic regression analysis includes $37,535$ drivers with 48 predictors after the transformation. The full data are randomly partitioned into $K=50$ sub-datasets, each with sample size of about $750$. The logistic regression estimation and inference results are provided in Table~\ref{tab:application}, which shows the estimated coefficients, standard errors and $p$-values of 48 potential risk factors obtained by CMLE, META and MODAC. Recall that CMLE is the centralized MLE method, which reads in all $K$ data batches and fit one logistic regression. CMLE gives the exact solution of maximum likelihood estimate and thus serves as our gold standard for comparisons. In terms of time, MODAC requires 0.66 seconds, one half of that by CMLE, which is 1.17 seconds. MODAC yields the exact same inference result as that of CMLE. Although META is the fastest and finishes in 0.03 seconds, its inference results deviate from those of CMLE and MODAC. For example, as inferred by both CMLE and MODAC, African American is less likely to have moderate to severe injury in a crash than White, and accidents are more likely to result in minor injuries on Wednesday than Sunday; in contrast, META is unable to capture these factors at the same confidence level.

\section{Discussion } \label{sec:discuss}
In this paper, we proposed a scalable regression method in the context of GLM with reliable statistical inference through the seminal work of confidence distribution. Although the divide-and-combine idea has been widely adopted in practice to solve computational challenges arising from the analysis of big data, statistical inference has been little investigated in such setting. We found in this paper that regularized estimation is appealing in the context of the GLM, especially in the logistic regression because regularization can effectively increase the numerical stability of regression analysis where there are many noisy features. In fact, such divide-and-combine inference may adopt other regularized estimators with regular limiting distributions, but we recommend sparse estimators for better numerical stability in estimating the bias terms and approximating the Fisher information matrices.

In practice, heterogeneity in covariate distributions may arise from various forms of distributed data storage over time and/or space. Some careful analyses are required to understand the nature of heterogeneity, which guide us to choose suitable methods in the integrative inference. Our method is essentially applicable to the targeted regression parameters that are the same across the sub-datasets, while both untargeted regression parameters and parameters of the second moments are allowed to differ across sub-datasets. When such targeted parameters are not clearly defined {\em a prior}, we may run an additional subgrouping analysis to identify the unknown subpopulations (see examples considered in \citep{tang2016fused,wang2016fused}), and then apply the proposed method to perform a group-based inference. Additionally, extension to allow unbalanced covariates' distributions and/or missing covariates across sub-datasets is an important direction to account for potential imbalances of data divisions, yet proper inference will require additional conditions similar to those proposed in \citep{li2018balancing, wang2015merging} to handle these complications.

Our method can be readily built in into some of the most popular open source parallel computing platforms, such as MapReduce \citep{dean2008mapreduce} and Spark \citep{zaharia2010spark},
to handle massive datasets where sample sizes are in the order of millions. Examples include estimating conversion rates using the Criteo online advertising data that have more than 2 million observations \citep{DiemertMeynet2017} and predicting patient disease status based on 9 million patients' electronic health records  \citep{wang2019fast}.
Although divide-and-combine is not needed for small datasets, our simulation results show that it is still preferable to impose regularization for large $p$ using the bias-correction technique.
For reproducibility, R code is provided in the Supplementary Material.

\section*{Acknowledgments}
The authors thank Editor-in-Chief, Associate Editor and two anonymous reviewers for their constructive comments. Zhou's research was partially supported by the Chinese Fundamental Research Funds for the Central Universities. Song's research was supported by an National Institutes of Health grant R01 ES024732 and an National Science Foundation grant DMS1811734.

\section*{Appendix Proofs}

\begin{proof}[\unskip\nopunct]
	{\em \textbf{Proof of Theorem~\ref{thm:dist}.}} We present here the key steps in the proof of Theorem~\ref{thm:dist} and relegate the complete proof to the Supplementary Material (Section~\ref{sec:supp:thm1proof}). We explicitly write subscript $k$ in the proof because the results will be used in Theorem~\ref{thm:cor2}. Denote some positive constants by $\mathcal{C}_l, l\in \{1,\dots,4\}$.
	For any fixed integer $q$, consider the bias-corrected estimator $\hat{\bgamma}_{\lambda_k,k} = \bH\hat{\bbeta}^c_{\lambda_k,k}$
	with $\hat\bbeta^c_{\lambda_k,k}  =  \hat{\bbeta}_{\lambda_k,k} + \{-\dot{\bS}_{n_k}(\hat{\bbeta}_{\lambda_k,k}) \}^{-1}\bS_{n_k}(\hat\bbeta_{\lambda_k,k})$, where the lasso estimator $\hat{\bbeta}_{\lambda_k,k}$ satisfies the KKT condition $\bS_{n_k}(\hat{\bbeta}_{\lambda_k,k})  -  \lambda_k \hat{\bkappa}_k = 0$.
	Let $\bA_{n_k}(\beta) = \frac{1}{n_k}\sum_{i=1}^{n_k}v(\mu_{k, i})\bx_{k,i}\bx_{k,i}^{T}$ and $\bB_{n_k} = \frac{1}{n_k}\sum_{i=1}^{n_k}\bx_{k,i}\bx_{k,i}^{T}$, where $\mu_{k, i} = g^{-1}(\bx^{T}_{k, i}\bbeta)$. Under condition (C2), it is easy to show that for any $\mu_{k, i} \in \Omega_{\delta}$,
	\begin{eqnarray} \label{eq:maineq}
	\textstyle
	c_kb_k^2\bI_{p \times p} \preceq c_k\bB_{n_k} \preceq \bA_{n_k}(\bbeta) \preceq C_k\bB_{n_k} \preceq C_kB_k^2\bI_{p\times p},
	\end{eqnarray}
	where $\bA \preceq \bB$ indicates $\bB-\bA$ is positive semi-definite. So $\dot{\bS}_{n_k}(\bbeta_{0,k}) = \phi^{-1}_k\bA_{n_k}(\bbeta_0)$ is invertible.
	With $PL(\hat{\bbeta}_{\lambda}; \bY, \bX) \geq PL(\bbeta_0; \bY, \bX)$, we have
	\begin{eqnarray*}
		\lambda_k \|\bbeta_{0,k}\|_1 &\geq& \frac{1}{n_k}\left\{\mathcal{L}_{n_k}(\bbeta_{0,k}; \bY_k, \bX_k) - \mathcal{L}_{n_k}(\hat{\bbeta}_{\lambda_k,k}; \bY_k, \bX_k)\right\} +\lambda_k\|\hat{\bbeta}_{\lambda_k,k}\|_1\\
		&=&-\bS_{n_k}(\bbeta_{0,k})^{T}(\hat{\bbeta}_{\lambda_k,k} - \bbeta_{0,k}) +
		\frac{1}{2\hat{\phi}_k} (\hat{\bbeta}_{\lambda_k,k} - \bbeta_{0,k})^{T}\bA_{n_k}(\tilde{\bbeta}_k)(\hat{\bbeta}_{\lambda_k,k} - \bbeta_{0,k})
		+\lambda_k\|\hat{\bbeta}_{\lambda_k,k}\|_1,
	\end{eqnarray*}
	where $\tilde{\bbeta}_k$ is a certain value between $\bbeta_{0,k}$ and $\hat{\bbeta}_{\lambda_k,k}$. It follows that
	\begin{equation*}
		\|\bP^{1/2}_{n_k}(\tilde{\bbeta}_k)\bX_k(\hat{\bbeta}_{\lambda_k,k} - \bbeta_{0,k})\|_2^2/(n_k\hat{\phi}_k) + 2\lambda_k\|\hat{\bbeta}_{\lambda_k,k}\|_1
		\leq 2\bS_{n_k}(\bbeta_{0,k})^{T}(\hat{\bbeta}_{\lambda_k,k} - \bbeta_{0,k}) + 2\lambda_k\|\bbeta_{0,k}\|_1.
	\end{equation*}
	According to Corollary 6.2 in \cite{buhlmann2011statistics} and conditions (C1)-(C3), we show that
	\begin{eqnarray}\label{eq:laa}
	\|\bX_k(\hat{\bbeta}_{\lambda_k,k} - \bbeta_{0,k})\|_2^2/{n_k} + \lambda_k\|\hat{\bbeta}_{\lambda_k,k} - \bbeta_{0,k}\|_1 \leq \mathcal{C}_1\lambda_k^2 s_{0,k},
	\end{eqnarray}
	where $s_{0,k} = \sum_{j=1}^pI(\beta_{0,k,j}\neq 0)$.
	
	To show the consistency and asymptotic normality of $\hat{\bgamma}_{\lambda_k,k}$, we begin with the first-order Taylor expansion on the KKT condition. Under conditions (C1)-(C3), we obtain
	\begin{equation} \label{pf:e}
	\hat{\bgamma}_{\lambda_k,k} - \bgamma_{0,k} =  \hat{\phi}_k \bH \bA^{-1}_{n_k}(\bbeta_{0,k})\bS_{n_k}(\bbeta_{0,k})
	-\bR_{n_k}(\tilde{\bbeta}_k, \bbeta_{0,k}; \bH) + \bB_{n_k}(\hat{\bbeta}_{\lambda_k,k}, \bbeta_{0,k}; \bH),
	\end{equation}
	where $\bR_{n_k}(\tilde{\bbeta}_k, \bbeta_{0,k}; \bH) = \bH \bA^{-1}_{n_k}(\bbeta_{0,k})\frac{1}{n_k}\bX_k^{T}\left\{\bP_{n_k}(\tilde{\bbeta}_k) - \bP_{n_k}(\bbeta_{0,k}) \right\} \bX_k(\hat{\bbeta}_{\lambda_k,k} - \bbeta_{0,k})$ \normalsize
	and
	\begin{equation*}
	\begin{split}
			\bB_{n_k}(\hat{\bbeta}_{\lambda_k,k}, \bbeta_{0,k}; \bH) 
	= \  & \bH\hat\phi_k\left\{\bA^{-1}_{n_k}(\hat\bbeta_{\lambda_k,k}) -  \bA^{-1}_{n_k}(\bbeta_{0,k})\right\}\left\{\bS_{n_k}(\hat\bbeta_{\lambda_k,k}) - \bS_{n_k}(\bbeta_{0,k})\right\}  \\
	&+  \bH\hat\phi_k\left\{\bA^{-1}_{n_k}(\hat\bbeta_{\lambda_k,k}) -  \bA^{-1}_{n_k}(\bbeta_{0,k})\right\}\bS_{n_k}(\bbeta_{0,k})
	\overset{def}{=} \bI_{1,k} + \bI_{2,k}.
	\end{split}
	\end{equation*}
	Note that from condition (C2) and (\ref{eq:maineq}),
	\begin{equation} \label{eq:re}
	\|\bR_{n_k}(\tilde{\bbeta}_k, \bbeta_{0,k}; \bH)\|_2^2
	\leq \textstyle \mathcal{C}_2 c_k^{-2}b_k^{-4}p\left\{n_k^{-1}\|\bX_k(\hat{\bbeta}_{\lambda_k,k} - \bbeta_{0,k})\|_2^2\right\}^2 \overline{\sigma}^2(\bH)q.
	\end{equation}
	Similarly, we have
	\begin{eqnarray} \label{eq:bias}
	\|\bI_{1,k}\|_2^2 &\leq& \mathcal{C}_3p\left\{n_k^{-1}\|\bX_k(\hat{\bbeta}_{\lambda_k,k} - \bbeta_{0,k})\|_2^2\right\}^2\left(\lambda_k^2s_{0,k}^2 + \lambda_k s_{0,k} + 1\right)\overline{\sigma}^2(\bH)q,\nonumber\\
	\|\bI_{2,k}\|_2^2 &\leq& \mathcal{C}_4n_k^{-1}\lambda_k^2s_{0,k}^2\overline{\sigma}^2(\bH)q.
	\end{eqnarray}
	{\color{black} Furthermore, by the multivariate Lindeberg-Levy central limit theorem \citep{serfling2009approximation} and Slutsky's theorem, the first term in \eqref{pf:e} satisfies $\hat{\phi}_k\bH \bA^{-1}_{n,k}(\bbeta_{0,k})\bS_{n_k}(\bbeta_{0,k}) \overset{d}{\rightarrow} \mathcal{N}\left(0, \bJ_{\gamma}(\bbeta_{0,k})\right)$ asymptotically as $n_k \to \infty$. Also, under condition (C3) that $\lambda_k = O\left\{(\log p/n_k)^{1/2}\right\}$ and $s_{0,k} = o\left\{n_k^{\frac{1-\delta}{2}}(\log p)^{-1/2} \right\}$, inequalities (\ref{eq:laa}), (\ref{eq:re}) and (\ref{eq:bias}) guarantee $\|\bR_{n_k}(\tilde{\bbeta}_k, \bbeta_{0,k}; \bH)\|_{2} = o_p(n_k^{-1/2})$ and $ \|\bB_{n_k}(\hat{\bbeta}_{\lambda_k,k}, \bbeta_{0,k}; \bH)\|_{2} = o_p(n_k^{-1/2})$. Thus, Theorem~\ref{thm:dist} follows.}
\end{proof}

\begin{proof}[\unskip\nopunct]
	 {\em \textbf{Proof of Corollary~\ref{thm:cor2}.}} See the Supplementary Material.
\end{proof}

\begin{proof}[\unskip\nopunct]
	{\em \textbf{Proof of Theorem~\ref{thm:eff}.}} Denote $\br_{N}(\bgamma) = \frac{1}{N}\sum_{k=1}^K\partial \log \hat{h}_{n_k}(\bgamma)/\partial \bgamma$ and $\br(\bgamma) = \underset{n_{\inf}\to \infty}{\lim}\br_{N}(\bgamma)$. It is easy to show $\br(\hat\bgamma_{dac}) \to \bm{0}$. On the other hand,
	\begin{eqnarray} \label{eq:rN}
	\br_{N}(\bgamma_0) &=& \textstyle -\frac{1}{N}\sum_{k=1}^Kn_k
	\left\{\bH  \dot{\bS}^{-1}_{n_k}(\hat{\bbeta}_{\lambda_k, k}) \bH^{T}\right\}^{-1}\left\{\bgamma_0 - \bH \hat{\bbeta}_{\lambda_k, k} +  \bH \dot{\bS}^{-1}_{n_k}(\hat{\bbeta}_{\lambda_k, k})\bS_{n_k}(\hat{\bbeta}_{\lambda_k, k})\right\} \nonumber\\
	&=& \textstyle \frac{1}{N}\sum_{k=1}^Kn_k\left\{\bH  \dot{\bS}^{-1}_{n_k}(\hat{\bbeta}_{\lambda_k, k}) \bH^{T}\right\}^{-1}\left\{\bH \dot{\bS}^{-1}_{n_k}(\hat{\bbeta}_{\lambda_k, k})\bS_{n_k}(\hat{\bbeta}_{\lambda_k, k}) + \bH(\bbeta_0 - \hat{\bbeta}_{\lambda_k, k})\right\}  \nonumber\\
	&=&\textstyle \frac{1}{N}\sum_{k=1}^K n_k \left\{\bH  \dot{\bS}^{-1}_{n_k}(\bbeta_0) \bH^{T}\right\}^{-1} \bH \dot{\bS}^{-1}_{n_k}(\bbeta_0)\bS_{n_k}(\bbeta_0) + o_p\left\{N^{-1}\left(\sum_{k=1}^{K}n_k^{1/2}\right)\right\},
	\end{eqnarray} \normalsize where the second equality holds under conditions (C1)--(C4). Then, by the law of large numbers,
	$\br(\bgamma_0) = \sum_{k=1}^{K}\frac{n_k}{N}$ $E\left[ \left\{\bH  \dot{\bS}^{-1}_{n_k}(\bbeta_0) \bH^{T}\right\}^{-1} \bH \dot{\bS}^{-1}_{n_k}(\bbeta_0)\bS_{n_k}(\bbeta_0)\right] = \bm{0}$, where the first equation follows from the condition that $K = o(N)$ and the second equation follows from condition that $E\left\{ \bH\dot{\bS}^{-1}_{n_k}(\bbeta_0)\bS_{n_k}(\bbeta_0) \right\} = \bm{0}$. Furthermore, we have $\dot{\br}(\bgamma_0) = \bJ^{-1}_{\bgamma}(
	\bbeta_{0})$, which is a negative-definite matrix given conditions (C1) and (C2). By combining this with $\br(\bgamma_0) = \bm{0}$ and  $\br(\hat{\bgamma}_{dac}) \to \bm{0}$, the consistency of $\hat{\bgamma}_{dac}$ follows.
	
	By simple algebra, we obtain
	\begin{eqnarray*}
		\hat{\bgamma}_{dac} &=& \textstyle \left[\sum_{k=1}^Kn_k\left\{-
		\bH \dot{\bS}^{-1}_{n_k}(\hat{\bbeta}_{\lambda_k, k})\bH^{T}
		\right\}^{-1}
		\right]^{-1}\left[\sum_{k=1}^Kn_k\left\{-
		\bH \dot{\bS}^{-1}_{n_k}(\hat{\bbeta}_{\lambda_k, k})\bH^{T}\right\}^{-1}
		\hat{\bgamma}_{\lambda_k, k} \right] \nonumber\\
		&=& \textstyle
		\left[\frac{1}{N}\sum_{k=1}^{K}n_k\left\{\bH\dot{\bS}^{-1}_{n_k}(\bbeta_0)\bH^{T} \right\}^{-1} \right]^{-1}
		\left[\frac{1}{N}\sum_{k=1}^{K}n_k\left\{\bH\dot{\bS}^{-1}_{n_k}(\bbeta_0)\bH^{T}\right\}^{-1} \hat{\bgamma}_{\lambda_k, k} \right] +
		O_p(N^{-1}K) + o_p(N^{-1/2}),
	\end{eqnarray*}
	\normalsize and $\text{var}(\hat{\bgamma}_{dac}) = N^{-1}\bJ_{dac, \bgamma}(\bbeta_0)$. Applying the condition that $K = O(N^{1/2 - \xi})$ with $\xi \in (0, 1/2]$ and the central limit theorem, we establish the asymptotic normal distribution of $\hat{\bgamma}_{dac}$.
	
	Finally, it suffices to show that $\hat{\bgamma}_{mle}$ has the same asymptotic distribution as $\hat{\bgamma}_{dac}$. By the definition of $\hat{\bgamma}_{mle}$ in Theorem~\ref{thm:eff}, we have  $\textstyle \hat{\bgamma}_{mle} - \bgamma_0 = -\bH \dot{\bS}^{-1}_{N}(\bbeta_0)\bS_N(\bbeta_0) + o_p(N^{-1/2}).$ The asymptotically equivalent efficiency claimed in Theorem~\ref{thm:eff} follows by the central limit theorem.
\end{proof}

%

\bibliographystyle{myjmva}
\bibliography{references}



\end{document}


%
%
%
%

\begin{center}
	{\Large Supplementary Material for ``Distributed Simultaneous Inference in Generalized Linear Models via Confidence Distribution''} \\ 
	\ \\
	Lu Tang, Ling Zhou, Peter X.-K. Song
\end{center}

\section{Additional Proofs}

\subsection{Proof of Theorem~1 with $p = O(n^{\delta})$ for $\delta \in [0,1)$}  \label{sec:supp:thm1proof}
Since the theorem pertains to the asymptotics for each sub-dataset, we suppress the subset index $k$ for ease of exposition. Denote some positive constants by $\mathcal{C}_l, l = 1,\dots, 4$. 

\begin{proof}
	For any fixed integer $q$, let $\bH = (h_{ij})_{q \times p}$ be a matrix of rank $q$ with bounded maximum singular value $\overline{\sigma}(\bH) < \infty$. Consider the bias-corrected estimator of $p$-dimensional $\beta_0$ at a tuning parameter $\lambda>0$,
	\begin{eqnarray}
	\label{eqn:beta_c_emp} \hat{\bgamma}_\lambda = \bH\hat{\bbeta}^c_{\lambda},
	\end{eqnarray}
	where $\hat\bbeta^c_\lambda  =  \hat{\bbeta}_\lambda + \{-\dot{\bS}_n(\hat{\bbeta}_\lambda) \}^{-1}\bS_n(\hat\bbeta_\lambda)$, and the lasso estimator $\hat{\bbeta}_{\lambda}$ satisfying the Karush-Kuhn-Tucker (KKT) condition $\bS_n(\hat{\bbeta}_\lambda)  -  \lambda \hat{\bkappa} = 0$.
	Recall that $\hat{\bkappa} = (\hat{\kappa}_1, \cdots, \hat{\kappa}_p)^{T}$ where $\max_j|\hat{\kappa}_j| \leq 1$, and $\hat{\kappa}_j = \operatorname{sign}(\hat{\beta}_{\lambda,j})$ if $\hat{\beta}_{\lambda,j} \neq 0$. Let $\bA_n(\bbeta) = \frac{1}{n}\sum_{i=1}^nv(\mu_i)\bx_i\bx_i^{T}$ and $\bB_n = \frac{1}{n}\sum_{i=1}^n\bx_i\bx_i^{T}$. Under condition (C2), it is easy to obtain that 
	\begin{eqnarray}
	\label{eq:maineq}
	cb^2\bI_{p \times p} \preceq c\bB_n \preceq \bA_n(\bbeta) \preceq C\bB_n \preceq CB^2\bI_{p\times p},
	\end{eqnarray}
	for any $\mu_i \in \Omega_{\delta}$, where $\bA \preceq \bB$ indicates that $\bB-\bA$ is a positive semi-definite matrix. Hence, $\dot{\bS}_n(\bbeta_0) = \phi^{-1}\bA_n(\bbeta_0)$ is invertible. 
	
	The penalized likelihood satisfies $PL(\hat{\bbeta}_{\lambda}; \bY, \bX) \geq PL(\bbeta_0; \bY, \bX)$ which leads to
	\begin{eqnarray*}
		\lambda \|\bbeta_0\|_1 &\geq& \frac{1}{n}\left\{\mathcal{L}_n(\bbeta_{0}; \bY, \bX) - \mathcal{L}_n(\hat{\bbeta}_{\lambda}; \bY, \bX)\right\} +\lambda\|\hat{\bbeta}_{\lambda}\|_1\\
		&=&-\bS_n(\bbeta_0)^{T}(\hat{\bbeta}_{\lambda} - \bbeta_0) + 0.5\times \hat{\phi}^{-1}(\hat{\bbeta}_{\lambda} - \bbeta_0)^{T}A_n(\tilde{\bbeta})(\hat{\bbeta}_{\lambda} - \bbeta_0)   +\lambda\|\hat{\bbeta}_{\lambda}\|_1,
	\end{eqnarray*}
	where $\tilde{\bbeta}$ is a certain value between $\bbeta_0$ and $\hat{\bbeta}_{\lambda}$. It follows that  
	$$\|\bP^{1/2}_n(\tilde{\bbeta})\bX(\hat{\bbeta}_{\lambda} - \bbeta_0)\|_2^2/(n\hat{\phi}) + 2\lambda\|\hat{\bbeta}_{\lambda}\|_1
	\leq 2\bS_{n}(\bbeta_0)^{T}(\hat{\bbeta}_{\lambda} - \bbeta_0) + 2\lambda\|\bbeta_0\|_1.
	$$
	According to Corollary 6.2 in \cite{buhlmann2011statistics} and conditions (C1)-(C3), it is easy to show that 
	\begin{eqnarray}\label{eq:laa}
	\|\bX(\hat{\bbeta}_{\lambda} - \bbeta_0)\|_2^2/n + \lambda\|\hat{\bbeta}_{\lambda} - \bbeta_0\|_1  \leq \mathcal{C}_1\lambda^2 s_0,
	\end{eqnarray}
	where $s_0 = \sum_{j=1}^pI(\beta_{0, j}\neq 0) = |\mathcal{A}_0|$.
	
	To show the consistency and asymptotic normality of $\hat{\bgamma}_{\lambda}$, we begin with the first-order Taylor expansion on the KKT condition. Under conditions (C1)-(C3), we obtain
	\begin{equation} \label{pf:e} 
	\hat{\bgamma}_{\lambda} - \bgamma_0 =  \hat{\phi}\bH\bP^{-1}_n(\bbeta_0)\bS_{n}(\bbeta_0) 
	-\bR_{n}(\tilde{\bbeta}, \bbeta_0; \bH) + \bB_n(\hat{\bbeta}_{\lambda}, \bbeta_0; \bH),
	\end{equation} 
	where the second and third terms are given by
	\begin{eqnarray*}
		\bR_{n}(\tilde{\bbeta}, \bbeta_0; \bH) &=& \bH\bA^{-1}_n(\bbeta_0)\frac{1}{n}\bX^{T}\left\{\bP_{n}(\tilde{\bbeta}) - \bP_{n}(\bbeta_0)
		\right\} \bX(\hat{\bbeta}_{\lambda} - \bbeta_0)\\
		&\overset{def}{=}&\bH\bA^{-1}_n(\bbeta_0)\frac{1}{n}\bX^{T}\bZ(\hat{\bbeta}_{\lambda}, \tilde{\bbeta}),\\
		\bB_n(\hat{\bbeta}_{\lambda}, \bbeta_0; \bH) &=& \bH\hat\phi\left\{\bA^{-1}_n(\hat\bbeta_{\lambda}) -  \bA^{-1}_n(\bbeta_0)\right\}\left\{\bS_n(\hat\bbeta_\lambda) - \bS_n(\bbeta_0)\right\}
		 + \bH\hat\phi\left\{\bA^{-1}_n(\hat\bbeta_{\lambda}) -  \bA^{-1}_n(\bbeta_0)\right\}\bS_n(\bbeta_0)\\
		&\overset{def}{=}&\bI_1 + \bI_2,
	\end{eqnarray*}
	where $\bZ(\bbeta, \bbeta')$ is an $n$-dimensional vector with the $i$th element being $Z_{i}(\bbeta, \bbeta') = \bx_i^{T}(\bbeta - \bbeta_0) \left[v(\mu_i') - v(\mu_i)\right]$ where $\mu_i' = g^{-1}(\bx_i^T\bbeta') $ and $\mu_i = g^{-1}(\bx_i^T\bbeta_0) $. 
	
	Note that 
	\begin{eqnarray}\label{eq:re} \textstyle
	\|\bR_{n}(\tilde{\bbeta}, \bbeta_0; \bH)\|_2^2 &=& \textstyle \text{tr}\left\{ \bH\bA^{-1}_n(\bbeta_0)\frac{1}{n}\bX^{T}
	\bZ(\hat{\bbeta}_{\lambda}, \tilde{\bbeta})\bZ^{T}(\hat{\bbeta}_{\lambda}, \tilde{\bbeta})X\frac{1}{n}\bA^{-1}_n(\bbeta_0)\bH^{T}
	\right\}\nonumber\\
	&\leq& \textstyle \left[\sum_{j=1}^p\left(\frac{1}{n}\sum_{i=1}^{n}x_{ij}Z_i(\hat{\bbeta}_{\lambda}, \tilde{\bbeta})\right)^2\right]
	\text{tr}\left[\bH\bA^{-1}_n(\bbeta_0)\bA^{-1}_n(\bbeta_0)\bH^{T}\right]\nonumber\\
	&\leq & \textstyle \mathcal{C}_2p\left\{n^{-1}\|\bX(\hat{\bbeta}_{\lambda} - \bbeta_0)\|_2^2\right\}^2 c^{-2}b^{-4}\text{tr}(\bH\bH^{T}) \nonumber\\
	&\leq& \textstyle \mathcal{C}_2c^{-2}b^{-4}p\left\{n^{-1}\|\bX(\hat{\bbeta}_{\lambda} - \bbeta_0)\|_2^2\right\}^2 \overline{\sigma}^2(\bH)q,
	\end{eqnarray} \normalsize
	where the second inequality follows from condition (C2) and the above inequality (\ref{eq:maineq}). Similarly, we have
	\begin{eqnarray} \textstyle
	\label{eq:bias}
	\|\bI_1\|_2^2 &=& \text{tr}\left[\bH \bA^{-1}_n(\hat{\bbeta}_{\lambda})\left\{\frac{1}{n}\bX^{T}\text{diag}\left[v\left\{g^{-1}(\bx_i^{T}\hat{\bbeta}_{\lambda})\right\} - v\left\{g^{-1}(\bx_i^{T}\bbeta_0)\right\}\right]\bX\right\}\bA^{-1}_n(\bbeta_0)\right.\nonumber\\
	&&\left.\times  \left\{\bA_n(\tilde{\bbeta}) - \bA_n(\bbeta_0) + \bA_{n}(\bbeta_0)\right\}(\hat{\bbeta}_{\lambda} - \bbeta_0)(\hat{\bbeta}_{\lambda} - \bbeta_0)^{T}
	\left\{\bA_n(\tilde{\bbeta}) - \bA_n(\bbeta_0) + \bA_{n}(\bbeta_0)\right\}\right.\nonumber\\
	&&\left. \times \bA_n(\bbeta_0)^{-1}\left\{\frac{1}{n}\bX^{T}\text{diag}\left[v\left\{g^{-1}(\bx_i^{T}\hat{\bbeta}_{\lambda})\right\} - v\left\{g^{-1}(\bx_i^{T}\bbeta_0)\right\}\right]\bX\right\}\bA^{-1}_n(\hat\bbeta_\lambda)\bH^{T}\right]\nonumber\\
	&\leq & \mathcal{C}_3p\left\{n^{-1}\|\bX(\hat{\bbeta}_{\lambda} - \bbeta_0)\|_2^2\right\}^2\left(\lambda^2s_0^2 + \lambda s_0 + 1\right)\overline{\sigma}^2(\bH)q,\nonumber\\
	\|\bI_2\|_2^2 &=& \text{tr}\left[\bH \bA^{-1}_n(\bbeta_0)\left\{\frac{1}{n}\bX^{T}\left[v\left\{g^{-1}(\bx_i^{T}\hat{\bbeta}_{\lambda})\right\} - v\left\{g^{-1}(\bx_i^{T}\bbeta_0)\right\}\right]\bX\right\}\bA^{-1}_n(\hat{\bbeta}_{\lambda})\bS_n(\bbeta_0)\right.\nonumber\\
	&&\left.\times  \bS^{T}_n(\bbeta_0) \bA^{-1}_n(\hat{\bbeta}_{\lambda})\left\{\frac{1}{n}\bX^{T}\text{diag}\left[v\left\{g^{-1}(\bx_i^{T}\hat{\bbeta}_{\lambda})\right\} - v\left\{g^{-1}(\bx_i^{T}\bbeta_0)\right\}\right]\bX\right\}\bA^{-1}_n(\bbeta_0)\bH^{T}\right]\nonumber\nonumber\\
	&\leq& \mathcal{C}_4n^{-1}\lambda^2s_0^2\overline{\sigma}^2(\bH)q,
	\end{eqnarray} \normalsize
	
	By using the central limit theorem and Slutsky's theorem, the first term in \eqref{pf:e} satisfies $\hat{\phi}\bH\bA^{-1}_n(\bbeta_0)\bS_{n}(\bbeta_0) \overset{d}{\rightarrow} \mathcal{N}\left(0, \bJ_{\bgamma}(\bbeta_0)\right)$ asymptotically as $n \to \infty$.
	Also, under condition (C3) that $\lambda = O\left\{(\log p/n)^{1/2}\right\}$ and $s_0 = o\left\{n^{\frac{1-\delta}{2}}(\log p)^{-1/2} \right\}$, inequalities (\ref{eq:laa}), (\ref{eq:re}) and (\ref{eq:bias}) guarantee $\|\bR_{n}(\tilde{\bbeta}, \bbeta_0; \bH)\|_{2} = \|\bB_{n}(\hat{\bbeta}_{\lambda}, \bbeta_0; \bH)\|_{2} = o_p(n^{-1/2})$. Thus, the proof of Theorem~1 is completed.
\end{proof}

\subsection{Proof of Corollary~1 for tuning parameter $\lambda$ selected by R-fold Cross-validation} \label{sec:supp:cor2proof}
Given any integer $R < \infty$, let $\bW_{r, train}$ and $\bW_{r, val}$ denote the $r$th training-and-validation partition of data $\bW = (\bY, \bX)$, for $r = 1, \dots, R$. Denote $\hat{\bbeta}_{\lambda_{cv}}(\bW)$ the estimator with the cross-validated tunning parameter $\lambda_{cv}$. To prove Theorem~1 for $\hat{\bbeta}_{\lambda_{cv}}(\bW)$, it is sufficient to show that the inequality (\ref{eq:laa}) holds for $\hat{\bbeta}_{\lambda_{cv}}(\bW)$; that is, $$\|\bX\left\{\hat{\bbeta}_{\lambda_{cv}}(\bW) - \bbeta_0\right\}\|_2^2/n + \lambda\|\hat{\bbeta}_{\lambda_{cv}}(\bW) - \bbeta_0\|_1  \leq \mathcal{C}_5\lambda^2 s_0,$$ where $\lambda = O\left\{(\log p)^{1/2}n^{-1/2}\right\}$. Following the definition of cross-validation procedure, we have that 
\begin{eqnarray}
0 &\leq& \sum_{r=1}^R\left[\mathcal{D}\left\{\hat{\bbeta}_{\lambda}(\bW_{r, train}); \bW_{r, val}\right\} - \mathcal{D}\left\{\hat{\bbeta}_{\lambda_{cv}}(\bW_{r, train}); \bW_{r, val}\right\}\right] \nonumber\\
&=&2\phi\sum_{r=1}^{R}\left[ \mathcal{L}\left\{\hat{\bbeta}_{\lambda_{cv}}(\bW_{r, train}); \bW_{r, val}\right\} - \mathcal{L}\left\{\hat{\bbeta}_{\lambda}(\bW_{r, train}); \bW_{r, val}\right\}\right] \nonumber\\
&=&2\phi\sum_{r=1}^{R}\left[ \mathcal{L}(\bbeta_0; \bW_{r, val}) + \bS(\bbeta_0; \bW_{r, val})^{T}\left\{\hat{\bbeta}_{\lambda_{cv}}(\bW_{r, train}) - \bbeta_0\right\}\right.\nonumber\\
&&\hspace{1.3cm}\left. + \frac{1}{2}\left\{\hat{\bbeta}_{\lambda_{cv}}(\bW_{r, train}) - \bbeta_0\right\}^{T}
\left\{\dot{\bS}({\bbeta}_{1, r}; \bW_{r, val})\right\}\left\{\hat{\bbeta}_{\lambda_{cv}}(\bW_{r, train}) - \bbeta_0\right\}\right] \nonumber\\
&&-2\phi\sum_{r=1}^{R}\left[ \mathcal{L}(\bbeta_0; \bW_{r, val}) + \bS(\bbeta_0; \bW_{r, val})^{T}\left\{\hat{\bbeta}_{\lambda}(\bW_{r, train}) - \bbeta_0\right\}\right.\nonumber\\
&&\hspace{1.3cm}\left. + \frac{1}{2}\left\{\hat{\bbeta}_{\lambda}(\bW_{r, train}) - \bbeta_0\right\}^{T}
\left\{\dot{\bS}({\bbeta}_{2, r}; \bW_{r, val})\right\}\left\{\hat{\bbeta}_{\lambda}(\bW_{r, train}) - \bbeta_0\right\}\right]\nonumber,
\end{eqnarray}
where $\hat{\bbeta}_{\lambda}(W)$ is the estimator with $\lambda$, $\mathcal{D}(\bbeta; \bW)$, $\mathcal{L}(\bbeta; \bW)$, $\bS(\bbeta; \bW)$ are the deviance function, log-likelihood function, and score function based on data $\bW$ and parameters $\bbeta$, respectively, and additionally, $\bbeta_{1, r}$ is a value between $\bbeta_0$ and $\hat{\bbeta}_{\lambda_{cv}}(\bW_{r, train})$, and ${\bbeta}_{2, r}$ is a value between $\bbeta_0$ and $\hat{\bbeta}_{\lambda}(\bW_{r, train})$, for $r = 1, \dots, R$.

Given the fact that $\bS(\bbeta_0; \bW_{r, val})$ and $\hat{\bbeta}_{\lambda_{cv}}(\bW_{r, train})$ $( \text{or} \hspace{0.2cm} \hat{\bbeta}_{\lambda}(\bW_{r, train}))$ are based on independent datasets $\bW_{r, val}$ and $\bW_{r, train}$, it is easy to show that the expectations of quantities $\bS(\bbeta_0; \bW_{r, val})^{T}\left\{\hat{\bbeta}_{\lambda_{cv}}(\bW_{r, train}) - \bbeta_0\right\}$ and $\bS(\bbeta_0; \bW_{r, val})^{T}\left\{\hat{\bbeta}_{\lambda}(\bW_{r, train}) - \bbeta_0\right\}$ are zero. Thus, the above inequality implies that
\begin{eqnarray*}
	&&\sum_{r=1}^R\left\{\hat{\bbeta}_{\lambda_{cv}}(\bW_{r, train}) - \bbeta_0\right\}^{T}
	\left\{\frac{1}{n_{r}}\bX^T_{r, val}\bP_{n_{r}}(\bbeta_{1, r})\bX_{r, val}\right\}\left\{\hat{\bbeta}_{\lambda_{cv}}(\bW_{r, train}) - \bbeta_0\right\} \nonumber\\
	&&\leq \sum_{r=1}^R\left\{\hat{\bbeta}_{\lambda}(\bW_{r, train}) - \bbeta_0\right\}^{T}
	\left\{\frac{1}{n_{r}}\bX^T_{r, val}\bP_{n_{r}}(\bbeta_{2, r})\bX_{r, val}\right\}\left\{\hat{\bbeta}_{\lambda}(\bW_{r, train}) - \bbeta_0\right\},
\end{eqnarray*}
where $n_{r}$, $\bX_{r, val}$ and $\bP_{n_{r}}(\bbeta)$ are the sample size, the set of covariates, and the diagonal weight matrix given $\bbeta$, of the $r$th validation data. Applying conditions (C1) and (C2), we obtain 
\begin{eqnarray}
\label{eq:l2norm}
\sum_{r=1}^R\|\bX_{r, val}\left\{\hat{\bbeta}_{\lambda_{cv}}(\bW_{r, train}) - \bbeta_0\right\}\|_2^2/n_{r} \leq \mathcal{C}_6R\lambda^2s_0, 
\end{eqnarray}
provided that $\frac{n - n_{r}}{n} = O(1)$, for $r = 1, \dots, R$.

On the other hand, it is easy to show that
\begin{eqnarray}
\label{eq:l1norm}
\|n_{r}^{-1/2}\bX_{r, val}\left\{\hat{\bbeta}_{\lambda_{cv}}(\bW_{r, train}) - \bbeta_0\right\}\|_1 \leq n_{r}^{1/2}\|n_r^{-1/2}\bX_{r, val}\left\{\hat{\bbeta}_{\lambda_{cv}}(\bW_{r, train}) - \bbeta_0\right\}\|_2 .
\end{eqnarray}

Combining the inequality (\ref{eq:l1norm}) with (\ref{eq:l2norm}), we have $\|n_{r}^{-1}\bX_{r, val}\left\{\hat{\bbeta}_{\lambda_{cv}}(\bW_{r, train}) - \bbeta_0\right\}\|_1 \leq \mathcal{C}_7R\lambda\sqrt{s_0}.$ Then, it follows that $ \|\hat{\bbeta}_{\lambda_{cv}}(\bW_{r, train}) - \bbeta_0\|_1 \leq \mathcal{C}_8R\lambda\sqrt{s_0}$ by condition (C2), 
$\|\bX\|_{\infty} = O(1)$. Moreover, since $n^{-1/2}\bX$ has bounded lower and upper singular values, we have
$$\|\bX_{r, train}\left\{\hat{\bbeta}_{\lambda_{cv}}(\bW_{r, train}) - \bbeta_0\right\}\|_2^2/(n-n_r) + \lambda\|\hat{\bbeta}_{\lambda_{cv}}(\bW_{r, train}) - \bbeta_0\|_1  \leq \mathcal{C}_9\lambda^2 s_0,$$
which is the result given by \eqref{eq:laa} in the proof of Theorem~1. 
Then the remainder of the proof follows similar steps to Theorem~1. So, we prove that Theorem~1 holds for estimator $\hat{\gamma}_{\lambda_{cv}}$ with $\lambda_{cv}$ being selected via the R-fold cross-validation procedure.

\section{Additional Simulation Results for Section~4}

\subsection{Sensitivity of $\omega$ in Majority Voting} \label{sec:supp:omega}

\begin{figure}
	\centering
	\begin{subfigure}[b]{0.32\textwidth}
		\centering
		\caption{Linear}
		\label{fig:votinga}		
		\includegraphics[width=\textwidth]{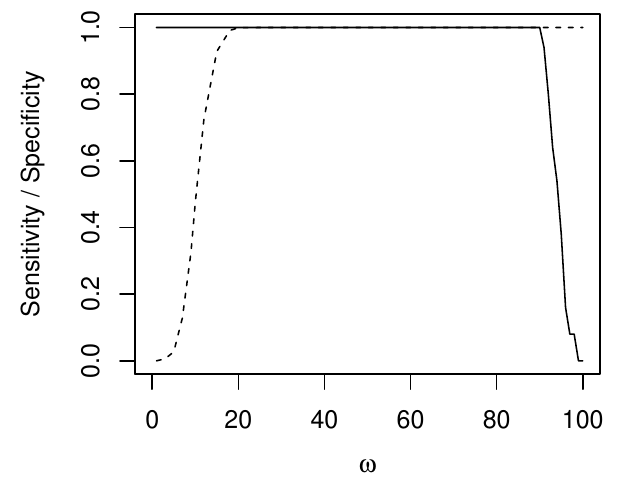}
	\end{subfigure}
	\begin{subfigure}[b]{0.32\textwidth}
		\centering
		\caption{Logistic}	
		\label{fig:votingb}		
		\includegraphics[width=\textwidth]{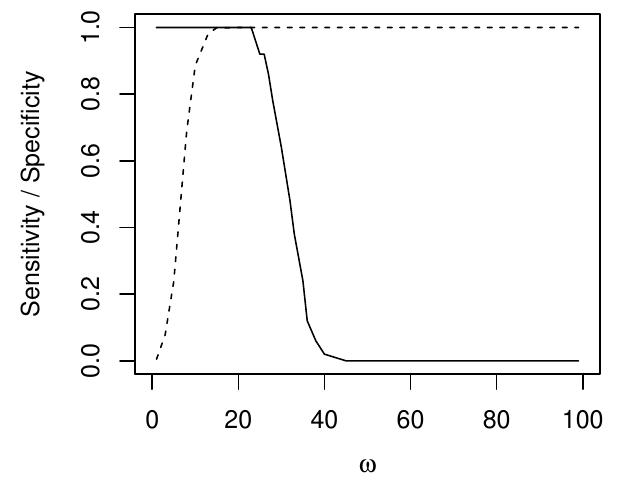}
	\end{subfigure}
	\begin{subfigure}[b]{0.32\textwidth}
		\centering
		\caption{Poisson}	
		\label{fig:votingc}		
		\includegraphics[width=\textwidth]{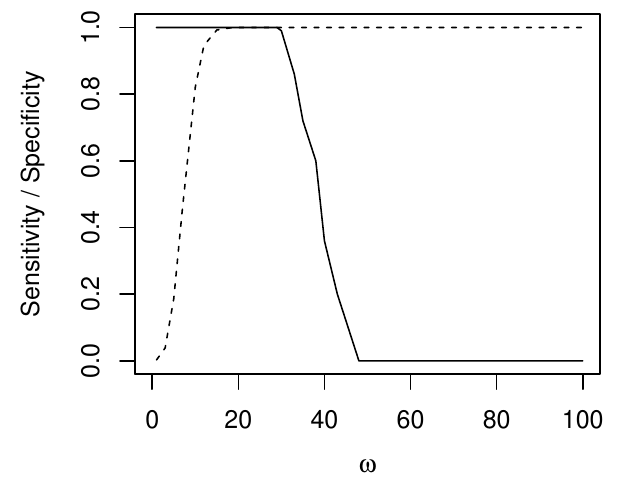}
	\end{subfigure}
	\caption{Sensitivity (solid) and specificity (doted) of MV as voting threshold $\omega$ varies from 0 to 100. The total sample size $N=50,000$, the number of split $K=100$, and the number of covariates $p=300$.}
	\label{fig:voting}
\end{figure}

Figure \ref{fig:voting} presents a sensitivity analysis of variable selection performance of the MV method by \cite{chen2014split} with respect to the choice of $\omega$ under three models, linear, logistic and Poisson. We consider $N=50,000$, $p=300$ and $s_0=10$. We vary the number of subsets $K$ and the correlation coefficient $\rho$ from a compound symmetric structure. The non-zero coefficients are set to $0.3$ for linear models, $0.3$ for logistic models, and $0.1$ for Poisson models. As shown, clearly the linear model is much more robust than the other two models by allowing a much wider range of $\omega$ to achieve the highest sensitivity and specificity. However, for logistic and Poisson models, only a very small range of $\omega$ around 20 is optimal for variable selection. The performance out of such ranges drops quickly. This poses a potential issue to real data analysis when the best range of $\omega$ is unknown. 

\subsection{Covariate Correlation versus Coverage Probability} \label{sec:supp:correlated}

To establish some guidelines about how to select $n_k$, we consider an additional simulation in which the correlation between covariates varies in terms of correlation coefficients $\rho$, and evaluate the performance of MODAC and META under different choices of $K$. Table~\ref{tab:rho_change} provides statistical inference results. The asymptotic confidence intervals of $\bbeta_{\mathcal{A}_0}$ of MODAC achieve the 95\% nominal coverage in most scenarios, except for the logistic regression with $\rho$ being small. Clearly, better performance of coverage occurs with bigger sub-dataset sizes. It is interesting to see that the performance gets better when the correlation $\rho$ gets larger. The poorer performance of MODAC in the logistic regression with a small $\rho$ may be due to the curse of dimensionality. As pointed out by \cite{hall2005geometric}, data tend to lie deterministically at the vertices of a regular simplex when the number of independent covariates goes to infinity and sample size is fixed. In other words, a limited amount of data would be problematic to make a valid statistical inference. On the other hand, larger correlation $\rho$ reduces effective degrees of freedom which make statistical inference a relatively easier task. Overall, the coverage probabilities of MODAC are uniformly more consistent than those of META. Based on the empirical results of MODAC, in practice, we suggest choosing a reasonably large $n_k$ in the logistic regression when covariates have weak dependence.

\begin{table}
	\caption{Simulation results when $N=50,000$ and $p=300$ for linear, logistic and Poisson models. Methods with different size of partition $K$ and  compound symmetric correlation $\rho$ are compared. $\mathcal{A}_0$ and $\mathcal{A}_0^c$ denote the set of non-zero and zero coefficients in $\bbeta_0$, respectively. Results are from an average of 100 replications. MODAC denotes our proposed method of divide-and-combine and META denotes the meta-analysis method. }
	\label{tab:rho_change}
	\resizebox{\columnwidth}{!}{
		\begin{tabular}{rrlcrrrrrrrr}
			\hline
			& & & & MODAC & MODAC & MODAC & MODAC & META & META & META & META\\
			$K$ & $n_k$ & Type & Set & $\rho=0$ & $\rho=0.3$ & $\rho=0.5$ & $\rho=0.8$ & $\rho=0$ & $\rho=0.3$ & $\rho=0.5$ & $\rho=0.8$ \\ 
			\hline
			50 & 1000 & Gaussian & $\mathcal{A}_0$ & 0.96 & 0.94 & 0.94 & 0.94 & 0.96 & 0.94 & 0.94 & 0.94 \\ 
			50 & 1000 & Binomial & $\mathcal{A}_0$ & 0.48 & 0.70 & 0.82 & 0.92 & 0.00 & 0.00 & 0.36 & 0.00 \\ 
			50 & 1000 & Poisson & $\mathcal{A}_0$ & 0.91 & 0.94 & 0.94 & 0.95 & 0.63 & 0.82 & 0.87 & 0.92 \\ 
			25 & 2000 & Gaussian & $\mathcal{A}_0$ & 0.97 & 0.96 & 0.96 & 0.95 & 0.97 & 0.96 & 0.96 & 0.95 \\ 
			25 & 2000 & Binomial & $\mathcal{A}_0$ & 0.77 & 0.86 & 0.87 & 0.94 & 0.00 & 0.01 & 0.04 & 0.36 \\ 
			25 & 2000 & Poisson & $\mathcal{A}_0$ & 0.92 & 0.96 & 0.95 & 0.95 & 0.79 & 0.88 & 0.92 & 0.93 \\ 
			10 & 5000 & Gaussian & $\mathcal{A}_0$ & 0.95 & 0.95 & 0.95 & 0.95 & 0.94 & 0.95 & 0.95 & 0.95 \\ 
			10 & 5000 & Binomial & $\mathcal{A}_0$ & 0.92 & 0.94 & 0.93 & 0.93 & 0.72 & 0.78 & 0.82 & 0.90 \\ 
			10 & 5000 & Poisson & $\mathcal{A}_0$ & 0.96 & 0.95 & 0.95 & 0.96 & 0.92 & 0.93 & 0.94 & 0.95 \\ [5pt]
			50 & 1000 & Gaussian & $\mathcal{A}_0^c$ & 0.95 & 0.95 & 0.95 & 0.95 & 0.95 & 0.95 & 0.95 & 0.95 \\ 
			50 & 1000 & Binomial & $\mathcal{A}_0^c$ & 0.96 & 0.96 & 0.96 & 0.95 & 1.00 & 1.00 & 0.94 & 0.18 \\ 
			50 & 1000 & Poisson & $\mathcal{A}_0^c$ & 0.95 & 0.95 & 0.95 & 0.95 & 0.92 & 0.92 & 0.93 & 0.93 \\ 
			25 & 2000 & Gaussian & $\mathcal{A}_0^c$ & 0.95 & 0.95 & 0.95 & 0.95 & 0.95 & 0.95 & 0.95 & 0.95 \\ 
			25 & 2000 & Binomial & $\mathcal{A}_0^c$ & 0.96 & 0.96 & 0.95 & 0.95 & 0.99 & 0.99 & 0.99 & 1.00 \\ 
			25 & 2000 & Poisson & $\mathcal{A}_0^c$ & 0.95 & 0.95 & 0.95 & 0.95 & 0.93 & 0.94 & 0.94 & 0.94 \\ 
			10 & 5000 & Gaussian & $\mathcal{A}_0^c$ & 0.95 & 0.95 & 0.95 & 0.95 & 0.95 & 0.95 & 0.95 & 0.95 \\ 
			10 & 5000 & Binomial & $\mathcal{A}_0^c$ & 0.95 & 0.95 & 0.95 & 0.95 & 0.97 & 0.97 & 0.97 & 0.97 \\ 
			10 & 5000 & Poisson & $\mathcal{A}_0^c$ & 0.95 & 0.95 & 0.95 & 0.95 & 0.94 & 0.95 & 0.95 & 0.95 \\  		
			\hline		
		\end{tabular}
	}
\end{table}

\subsection{Comparison with Speed GLM (SPGLM)}   \label{sec:supp:speedglm}

\begin{table}
	\caption{Simulation results of SPGLM in comparison with the proposed MODAC and the gold standard CMLE when $N=50,000$ and $p=300$ for linear, logistic and Poisson models. Results are from an average of 100 replications. }
	\label{tab:speedglm}	
	\resizebox{\columnwidth}{!}{
		\begin{tabular}{lrrrrrrrrr}
			\hline
			& \multicolumn{3}{c}{Linear Model} & \multicolumn{3}{c}{Logistic Model} & \multicolumn{3}{c}{Poisson Model} \\
			& CMLE & SPGLM & MODAC & CMLE & SPGLM & MODAC & CMLE & SPGLM & MODAC \\ 
			\hline		
			Sensitivity & 1.00 & 1.00 & 1.00 & 1.00 & 1.00 & 1.00 & 1.00 & 1.00 & 1.00 \\ 
			Specificity & 0.95 & 0.95 & 0.95 & 0.95 & 0.95 & 0.96 & 0.95 & 0.95 & 0.95 \\ 
			MSE of $\hat{\bbeta}_{\mathcal{A}_0}$ ($\times 100$) & 0.01 & 0.01 & 0.01 & 0.08 & 0.08 & 0.10 & 0.01 & 0.01 & 0.01 \\ 
			MSE of $\hat{\bbeta}_{\mathcal{A}_0^c}$ ($\times 100$) & 0.01 & 0.01 & 0.01 & 0.08 & 0.08 & 0.07 & 0.01 & 0.01 & 0.01 \\ 
			Absolute bias of $\hat{\bbeta}_{\mathcal{A}_0}$ & 0.01 & 0.01 & 0.01 & 0.02 & 0.02 & 0.02 & 0.01 & 0.01 & 0.01 \\ 
			Absolute bias of $\hat{\bbeta}_{\mathcal{A}_0^c}$ & 0.01 & 0.01 & 0.01 & 0.02 & 0.02 & 0.02 & 0.01 & 0.01 & 0.01 \\ 
			Cov. prob. of $\bbeta_{\mathcal{A}_0}$ & 0.95 & 0.95 & 0.95 & 0.95 & 0.96 & 0.92 & 0.95 & 0.95 & 0.95 \\ 
			Cov. prob. of $\bbeta_{\mathcal{A}_0^c}$ & 0.95 & 0.95 & 0.95 & 0.95 & 0.95 & 0.96 & 0.95 & 0.95 & 0.95 \\ 
			Asymp. st. err. of $\hat{\bbeta}_{\mathcal{A}_0}$ & 0.01 & 0.01 & 0.01 & 0.03 & 0.03 & 0.03 & 0.01 & 0.01 & 0.01 \\ 
			Asymp. st. err. of $\hat{\bbeta}_{\mathcal{A}_0^c}$ & 0.01 & 0.01 & 0.01 & 0.03 & 0.03 & 0.03 & 0.01 & 0.01 & 0.01 \\ 
			Computation time & 34.85 & 11.02 & 2.14 & 66.01 & 15.18 & 10.53 & 42.26 & 15.92 & 25.08 \\
			\hline
		\end{tabular}
	}
\end{table}

Under the same simulation setting as that in Section~4,  we compare MODAC with  SPGLM from an existing R package for big data with $N \gg p$, \verb|speedglm|. This package produces the exact solution as that of the gold standard CMLE for large data that exceed memory limits. The algorithm of SPGLM is similar to the mini-batch gradient descent, where individual data batches are stored in separate hard drives. See Table~\ref{tab:speedglm} for a comparison between CMLE, SPGLM and MODAC, when $N=50,000$ and $K=100$. In this case, SPGLM achieves faster computation time than MODAC only in the Poisson model due probably to the inefficient programming of the R package \verb|glmnet| for Poisson regression. But as $N$ continue increases, SPGLM slows down dramatically because it uses a pseudo parallel algorithm which requires iteratively reading from sub-datasets, whereas MODAC does not.  To our best knowledge, none of existing packages has considered the parallel version of GLM in large data sets without using sub-datasets iteratively when the dimension of the variable is not small.

\section{Data Information}   \label{sec:supp:realdata}

The National Highway Traffic Safety Administration research and data website (\url{https://www.nhtsa.gov/research-data}) lists all of its research projects and data. In this paper, we use data from the National Automotive Sampling System (NASS), which includes the Crashworthiness Data System and the General Estimates System as detailed in \url{https://www.nhtsa.gov/research-data/national-automotive-sampling-system-nass}. We focus on the Crashworthiness Data System, whose raw data files are organized by year and can be downloaded at \url{ftp://ftp.nhtsa.dot.gov/NASS/}. We also include a copy of the cleaned dataset used in our real data analysis as part of the supplementary material.

\section{Software Code} \label{sec:supp:code}
Python and R code of the method is available at \url{http://www.umich.edu/~songlab/software}.

%

\bibliographystyle{myjmva}
\bibliography{references}

